\documentclass[aps,jcp,preprint,citeautoscript,showpacs,longbibliography]{revtex4}

\usepackage{graphicx}
\usepackage{caption}
\usepackage{subcaption}
\usepackage{amsmath}
\usepackage{amssymb}
\usepackage{tabularx}
\usepackage{placeins}
\usepackage{units}
\usepackage[usenames, dvipsnames]{color}
\usepackage{hyperref}
\usepackage{tikz}
\usepackage{multirow}

\usepackage{dcolumn}
\usepackage{bm}

\begin{document}

\title{A new generation of effective core potentials from correlated calculations: 2nd row elements}

\author{M. Chandler Bennett$^{1,2}$, Guangming Wang$^{1}$, Abdulgani Annaberdiyev$^{1}$, Cody A. Melton$^{1,2}$, Luke Shulenburger$^{2}$, and Lubos Mitas$^1$}

\affiliation{
1) Department of Physics, North Carolina State University, Raleigh, North Carolina 27695-8202, USA \\
}
\affiliation{
2) Sandia National Laboratories, Albuquerque, New Mexico 87123, USA \\
}

\date{\today}


\begin{abstract}
Very recently, we have introduced 
correlation consistent effective core potentials (ccECPs)  
derived from many-body approaches
with the main target being their use in explicitly correlated methods, while still usable in mainstream approaches. The ccECPs are based on 
reproducing excitation energies for a subset of valence states, namely, achieving near-isospectrality between the original and pseudo Hamiltonians. 
In addition, binding curves of dimer molecules were used for refinement and overall improvement of transferability over a range of bond lengths. 
Here we apply similar ideas to the 2nd row elements and study several aspects of the constructions in order to find the high accuracy solutions within the 
chosen ccECP forms with $3s,3p$ valence space (Ne-core). 
Our new constructions exhibit accurate low-lying atomic excitations and equilibrium molecular bonds (on average within $\approx$ $0.03$ eV and $3$ m\AA), however, the errors for Al and Si oxide molecules at short bond lengths are notably larger for both ours and existing effective core potentials.
Assuming this limitation, our ccECPs show a systematic balance between the criteria of atomic spectra accuracy and transferability for molecular bonds.
In order to provide another option with much higher uniform accuracy, we also construct He-core ccECPs for the whole 2nd row with typical discrepancies of $\approx 0.01$ eV or smaller. 
\end{abstract}

\maketitle

\section{Introduction}
The effective core potential (ECP) approximation has been vital in electronic structure calculations for several decades.
An ECP is used to replace an atom's core (nucleus and tightly bound electrons) and is constructed to mimic the original core's effects on the valence electrons. 
This approximation provides important benefits such as the removal of large energy scales associated with the core electrons while simultaneously enabling the implicit incorporation of the core's relativistic effects.  It generates smoother charge densities around the nucleus which, for instance, allows the use of much smaller plane wave basis cutoff energies or less extensive gaussian basis sets.
Consequently, in many cases, the availability of accurate ECPs can be \emph{the} deciding factor for whether a calculation is even feasible or not.
Given that the scope and accuracy of many-body electronic structure methods are continually increasing (true for both stochastic and basis set approaches, see \cite{kolorenc2011, wagner2016, dubecky2016, needs2010, booth2013}), it is important that the errors of the ECP approximation be kept as small as possible or at least kept comparable to other systematic errors that may be present.

Historically, ECPs have generally been formulated within effective mean-field (single-particle) theories where the concept of core-valence partitioning follows rather naturally.
A number of approaches have been established in the Density Functional Theory (DFT) \cite{Hamann, bachelet1984, tm:prb1991, vanderbilt1990, Hamann2013, goedecker1996,goedecker2013}, 
see also review \cite{Pickett}.
For the review of Dirac-Fock energy consistent constructions
see \cite{DolgCao}. Although
generating valence-only Hamiltonians is more straightforward in an effective one-particle setting, the performance of such ECPs within many-body theories is not guaranteed and can lead to results of mixed quality \cite{Shulenburger:2013prb} which can require adjustments to make them reliable within these theories \cite{foyevtsova2014, nazarov2016}. 
In order to achieve more consistent and systematic accuracy within many-body theories, we have recently proposed a general framework and provided examples of a new generation of correlation consistent ECPs (ccECPs) \cite{Bennett:2017jcp} that aim to reach beyond the current status. 
The key principles included the use of many-body constructions and benchmarking to ascertain the quality of the ECPs from the outset.
The first component of our strategy was based on near-isospectrality, within a subspace of valence states, between the all-electron and ECP Hamiltonians. 
Furthermore, the ECP properties were studied not only for atoms but also for small molecules to probe for transferability in bonded situations in order to validate their overall quality. 

Remarkably, for the first row, we found that it was possible to derive ECPs that show significant improvements in various valence properties at the CCSD(T) level of theory when compared with existing ECP tables or the all-electron uncorrelated-core (UC) approximation where excitations out of the core are suppressed. 
Quite surprisingly this was achieved with simple, ``minimal", non-divergent ECP parameterizations, i.e., with very few gaussians per nonlocal channel.

In order to maintain the simplicity and direct connection of the ECP to the original Hamiltonian, the solution was formulated as an inverse problem. Namely, we searched for an effective Hamiltonian operator that reproduces a given set of many-body eigenenergies and eigenstates, outside the core region, as closely as possible. 
Not surprisingly, this can lead to an ill-conditioned problem (high sensitivity to inputs) and/or a complicated optimization landscape with many near-degenerate minima.
Consequently, such optimizations can be rather difficult. 
Such difficulties can be understood conceptually, for instance, if one considers the well-known example from the spectral theory that two operators with different domains can 
have the same spectrum, i.e., they can be exactly isospectral. 
This non-uniqueness suggests that the problem requires adequate and proper constraining in order to arrive at a desirable solution with a reasonable amount of effort.
For the B, C, N, O and S elements, we were able to accomplish just that and produced ECPs for these atoms that exceed the accuracy of existing tabulated ECPs and that have led to significant improvements in transferability.

We also aimed to construct ECPs that are simple and can be employed in a variety of electronic structure packages and methods, i.e., as widely applicable as possible.
To this end, we chose a well-established form \cite{Burkatzki:2007jcp,Lester:2001jcp} that is non-singular and behaves quadratically at the origin and is parameterized by a short sum of radial powers multiplied by gaussians. This form leads to one-particle orbitals that also behave quadratically at the origin which makes their expansion in gaussians more efficient. Additionally, the smooth form provides computational savings for quantum Monte Carlo by reducing fluctuations.

Another aspect of our recently constructed ECPs was to probe for the transferability in bonds both near and out of the equilibrium configurations, having in mind, for example, high-pressure applications in future. 
Therefore we monitored discrepancies of dimer binding curves including short bond lengths up to the dissociation limit, where the binding energy goes to zero for the compressed bond.  
Note that this limit is particularly challenging since it directly probes whether the ECP
correctly mimics the (missing) core charge at short
lengths.
In cases where larger errors were present, we included the dimer into the construction as an additional constraint which improved the quality of the resulting ECPs. 
Indeed, we found that the key molecular properties were more accurately reproduced in this way. 
As an independent probe of transferability, we then tested hydride and oxide molecules.
Further benchmarks and testing will come from future applications for systems where all-electron correlated calculations can be carried out. 

In the present work, we extend our previous collection of correlation-consistent ECPs to include the entire 2nd row of elements Na through Ar (following the convention that H and He comprise the zeroth row). 
For these elements, we have further improved and modified the methodology used in generating the ECPs which we describe in more detail in what follows.
The constructed ECPs that we present here are not meant to be definitive. 
It is clear that with more elaborate forms, say, with a larger number of free parameters, the accuracy might still improved. 
Another consideration is that different applications might require further refinement or even new constructions.
For example, alkaline elements have nominally large cores and one or two valence electrons.
Clearly, there are applications where this would not provide an accurate replacement of the all-electron Hamiltonian (high pressures) and the inclusion of the outermost shell from the core to the valence space might be needed. 
This implies that one needs more options to guarantee the accuracy in various physical or chemical applications. 
Therefore keeping track of updates and any new constructions, as well as maintaining benchmark sets is an important part of providing clear choices and validations for future use of ECPs.

The paper is organized as follows. 
In section II, we discuss the different components of a number of atomic energy gaps of a variety of tabulated 2nd row ECPs and how this affects their agreement with all-electron gaps at the many-body level as well as how this motivated the particular construction we use for the ECPs presented in this work. 
We outline this construction in section III.
In section IV, we describe the particular parameterization we use for our ECPs.
Section V outlines the scheme we followed to construct the ECPs' corresponding valence basis sets.
In section VI, we share our results and present our ECPs together with several existing ones for comparison and highlight the shortcomings of Ne-core ECPs in particular settings (short polar bonds) and therefore, in addition, present He-core ECPs for the entire 2nd row that significantly improve the accuracy of ECPs with Ne-core partitioning and serve also as benchmarks for comparisons.
In section VII, we give our concluding remarks.

\section{ECP Atomic Correlation Energies}\label{sec:correlations}

We have looked in a more quantitative manner at the contributions from valence-valence correlation energies as well as Hartree-Fock (HF) energies for a variety of atomic excitations among a testbed of tabulated ECPs.
This provides additional insights into how various components of the energy are modified and respond to ECP differences (form, parameterization, etc.) which ultimately enabled us to improve our ECP construction scheme and helped to improve their overall performance.
Specifically, for a number of tabulated ECPs, we have analyzed the errors (relative to relativistic all-electron results) of the HF and correlation contributions to the CCSD(T) spectral energies for the Si and P atoms and we have looked at how the errors for these energy components change between different ECPs.
Relativity in all-electron calculations was included through the 10th-order Douglas-Kroll-Hess Hamiltonian for both HF and CCSD(T) approaches.

\begin{figure}[ht!]
    \centering
    \caption{For the silicon atom, the spread of CCSD(T) valence-valence correlation errors ($\Delta$cVV) and spread of HF errors ($\Delta$HF) for various excitation energies from a variety of previously tabulated ECPs, in particular, BFD\cite{Burkatzki:2007jcp}, CRENBL\cite{crenbl}, SBKJC\cite{SBK}, STU\cite{STU} and TN-DF\cite{Trail:2005jcp}. }
\includegraphics[width=0.45\textwidth]{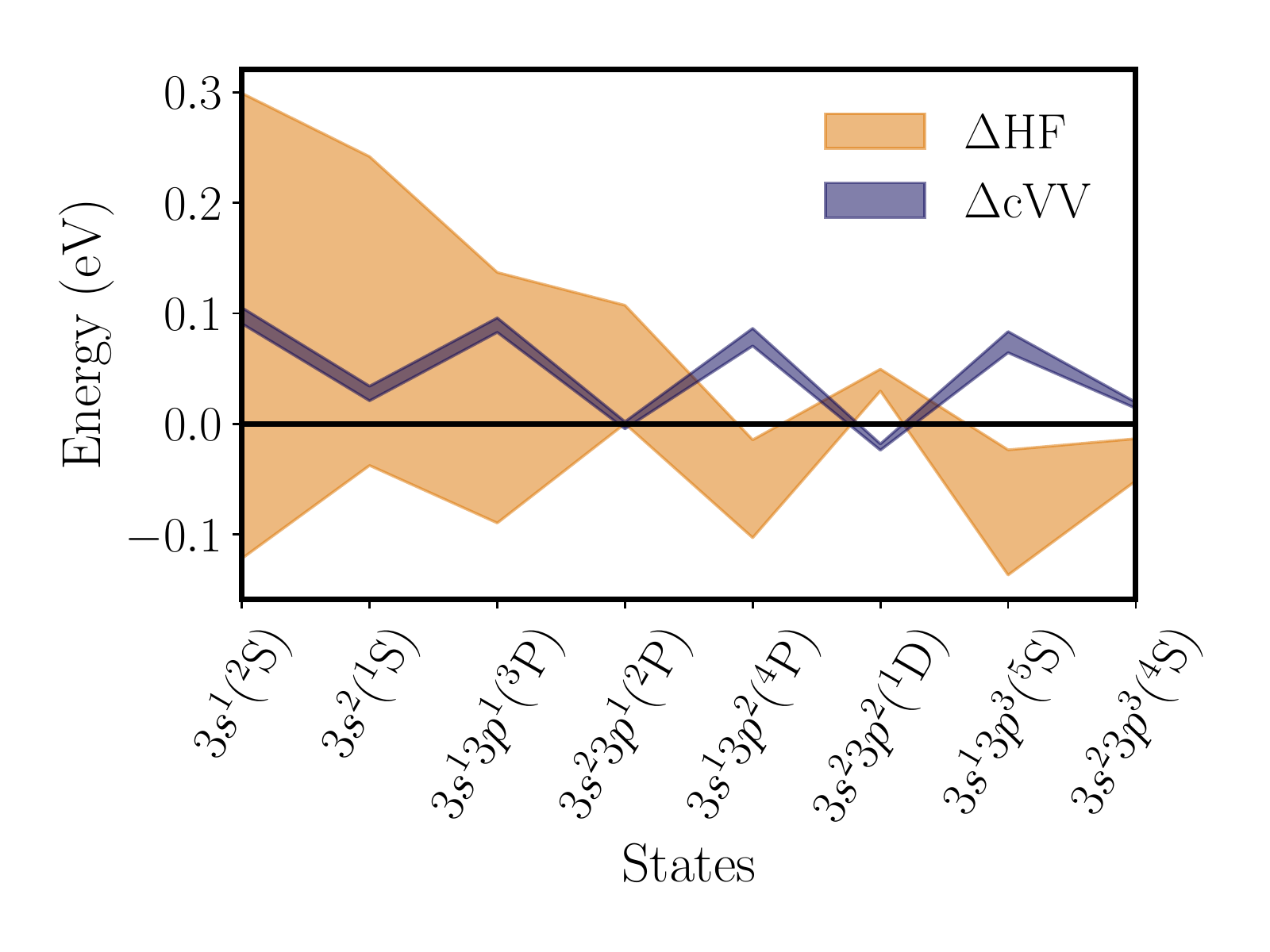}
\label{fig:si_dVV_dHF}
\centering
\end{figure}

\begin{figure}[ht!]
    \caption{For the phosphorus atom, the spread of CCSD(T) valence-valence correlation errors ($\Delta$cVV) and spread of HF errors ($\Delta$HF) for various excitation energies from a variety of previously tabulated ECPs, in particular, BFD\cite{Burkatzki:2007jcp}, CRENBL\cite{crenbl}, SBKJC\cite{SBK}, STU\cite{STU} and TN-DF\cite{Trail:2005jcp}.}
\includegraphics[width=0.45\textwidth]{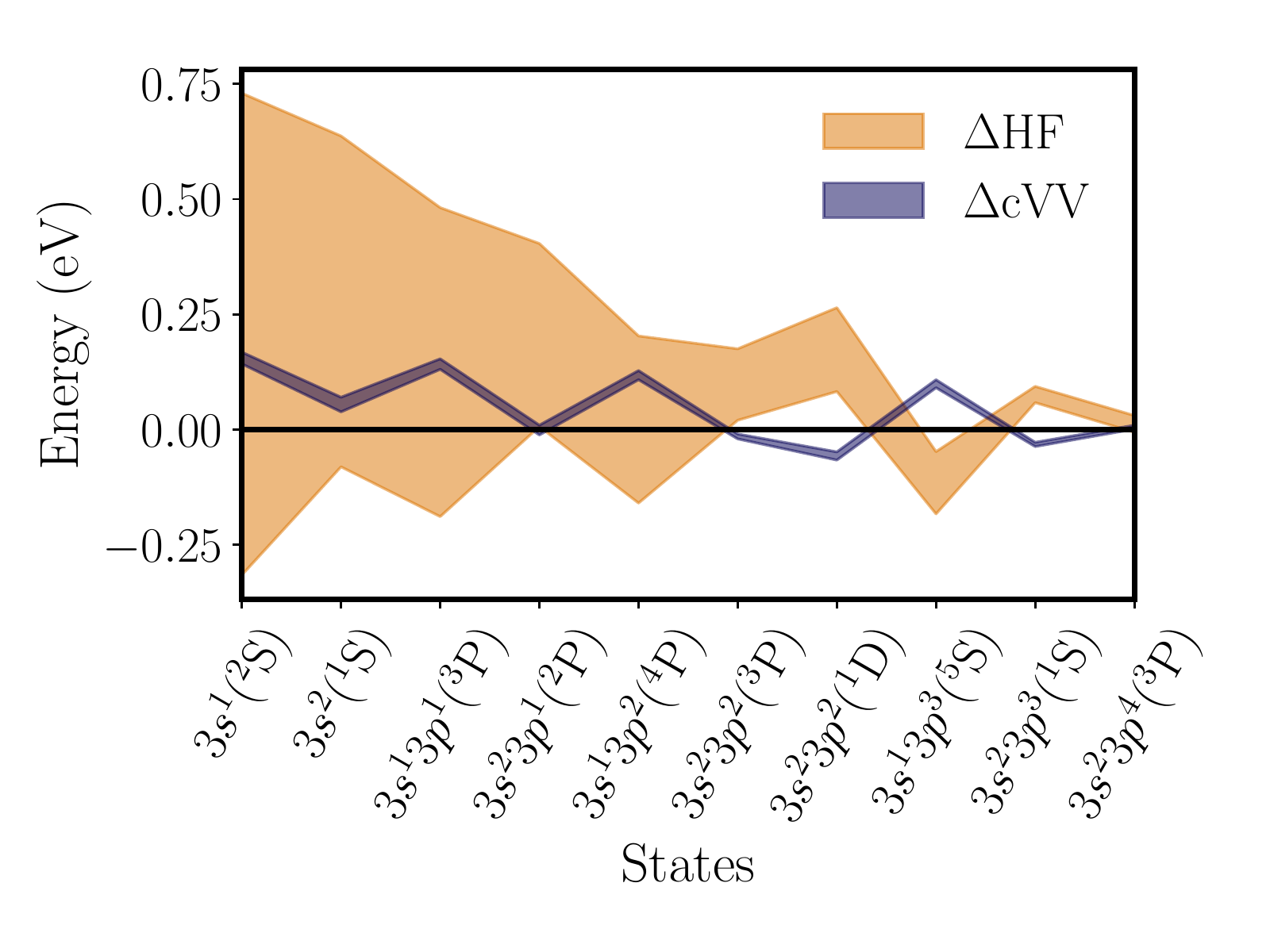}
\label{fig:p_dVV_dHF}
\centering
\end{figure}

In Figs. \ref{fig:si_dVV_dHF} and \ref{fig:p_dVV_dHF}, for Si and P, we show examples of the spread of HF and CCSD(T) valence-valence correlation errors across a number of previously tabulated ECPs, which employ a wide variety of parameterizations, for a number of excitation energies.
In particular, we plot the spread of the following quantities over the set of ECPs:
\begin{equation} \label{eq:hf-spread}
    \Delta {\rm HF}_s = \Delta E_s^{\rm ECP} - \Delta E_s^{\rm AE} \, ,
\end{equation}
where $\Delta E_s$ represents the energy gap between a given state and the ground state, and
\begin{equation} \label{eq:vv-spread}
	\Delta {\rm cVV}_s = \left|{\rm VV}_{{\rm corr,}s}^{ ECP }\right| - \left|{\rm VV}_{{\rm corr,}s}^{AE}\right| \, ,
\end{equation}
where ${\rm VV_{\rm corr}}$ is the valence-valence correlation energy.
In these figures, the spread (represented by the shaded areas) are bounded by the maximum discrepancy (upper bound) and minimum discrepancy (lower bound) among the various ECPs.
For the ECP cases, the valence-valence correlation is taken to be the fully correlated CCSD(T) energy of a particular state.
For the all-electron case, the valence-valence correlation is taken to be the UC CCSD(T) correlation energy of a particular state. 

There are several observations that can be gleaned from these plots. 
The first is that the ECPs' correlation energies are on average larger than the corresponding all-electron valence-only correlated values with the increase varying 
between 0 and 0.15 eV. 
This has been known for some time, see the analysis by Dolg \cite{Dolg:1996cpl}. 
The reason is the absence of the radial node(s) and increased smoothness of ECP (pseudo)orbitals since this increases the probability for unlike spin pairs of electrons to encounter each other, thus increasing the correlation energy. 
However, the next striking feature is the rigidity of the correlated energy differences across various ECPs, regardless of construction, whether it shows singular or bounded/smooth behavior of local and nonlocal terms at the origin or other details. 
Essentially the ECP correlation energies are almost invariant to the the form or the construction of ECP, typically within $0.02$ eV or less. 
Consequently, the largest differences between ECPs for atomic spectra come from fluctuations that originate in the HF component of energy differences. Finally, the third point is that aiming for atomic spectra accuracy better than 0.1 eV requires the HF component to ``compensate" for the rigid correlation contribution component for a number of states. 
Given this rigidity of ECP correlations, we incorporate mean-field data into the construction of the 2nd row ECPs along with many-body data in order to avoid large mean-field biases which we found helps to improve transferability (shown below in section V). In the following section we give the specifics of the procedure we use to construct the ECPs.

\section{Construction}\label{sec:construction}

The strategy that we employ to generate our ECPs is a combination of many-body energy consistency and single-body norm conservation. We provide the details of each and how they are combined to form our full objective function in this section.

\subsection{Many-body Energy Consistency}

In order to account for correlation when generating our ECPs, we have employed a spin-unrestricted CCSD(T) energy consistency scheme as part of our full objective function.
The details of this scheme are as follows: 
For a given atom, we first generate a large number of all-electron reference atomic states which includes various ionizations and spin-multiplicities. The states are calculated with the fully correlated (both core and valence electrons) spin-unrestricted CCSD(T) method. We account for relativity by using the 10th order Douglas-Kroll-Hess Hamiltonian \cite{Reiher:2004jcp}. 
We formulate this part of the objective function by calculating the same atomic states with a parameterized ECP at the same level of theory, and then vary the parameters until the gaps between these states and the neutral ground state agree with that from the all-electron to a satisfying level.
The discrepancies of the ECP gaps relative to the all-electron gaps are minimized in the least squares way which defines the first component of our objective function, $\mathcal{E}^2$, given by
\begin{equation} \label{eq:energy_soq}
    \mathcal{E}^2 = \sum_s \left( \Delta E_s^{\rm ECP } - \Delta E_s^{\rm AE} \right)^2,
\end{equation}
where $s$ labels a given atomic state and $\Delta E_s$ is the energy gap between the state and the neutral ground state and ``ECP" and ``AE" label the ECP and all-electron gaps, respectively.

The energetics from all-electron and ECP were calculated with the \textsc{Molpro} quantum chemistry package \cite{MOLPRO-WIREs}.
For all atoms, the uncontracted aug-cc-pCV5Z basis set \cite{Dunning:1989jcp}, where core state correlation functions are present, was utilized.
If linear dependencies developed as a result of uncontracting the basis, the problematic terms with nearly identical exponents were removed.
This same basis was used for calculating both the all-electron and ECP gaps in order to construct similar basis set errors between the two which would largely cancel when calculating their difference in Eq. \ref{eq:energy_soq}.
Quantitative estimates of the residual finite basis set errors were given in our previous work \cite{Bennett:2017jcp} for a test set of atoms from the 1st and 2nd row.

We chose not to use the $\mathcal{E}^2$ objective function exclusively when training our ECPs given that in doing so we observed an overall large increase in error of the HF gaps and subsequently poorer transferability when calculating potential energy surfaces of diatomic molecules (dimers and oxides) at the CCSD(T) level of theory. 
We suspect this is due to the rigidity in the ECP correlation energies as outlined in section \ref{sec:correlations}, namely, in order to reduce the total error of the CCSD(T) gap, the magnitude of the HF error increases to compensate the error from the more rigid correlations.
We observed that reaching small gap errors at the many-body level at the full expense of the HF gap errors showed tendencies to increase transferability errors. In fact, the best results came from a balance between the errors in correlation and HF components. In addition, significant 
variational freedom inside the core is proved to be helpful and 
we have therefore included a single-body norm conservation into our objective function, as outlined in the next subsection, in order to strike this balance.

\subsection{Single-body Norm Conservation}
{
In order to maintain accuracy at the HF level of theory, we employed a norm conservation scheme as another component of our full objective function.
We have taken a strategy similar to what has been used in previous works \cite{SBK,Lester:2001jcp,Christiansen:1979jcp,Trail:2005jcp}, however, the specifics of our approach are as follows.
For the Ne-core ECPs, we first generated all-electron scalar relativistic HF radial orbitals, $\phi^{\rm AE}_l$, where the 10th order DKH hamiltonian was employed. 
For each angular momentum channel, we chose a cutoff radius, $R_l$, corresponding to the outer most extremum of {$r^{\frac{4}{5}}\phi^{\rm AE}_l$}.
At this radius, we then obtained the orbital's norm inside this point, $N^{\rm AE}_l=\int_0^{R_l}\big(r^{l+1}\phi^{\rm AE}_l(r)\big)^2\mathrm{d}r$, its value, $V^{\rm AE}_l=\phi^{\rm AE}_l(R_l)$ and its slope, $S^{\rm AE}_l=\frac{\rm d}{\mathrm{d}r}\phi^{\rm AE}_l(r)\Big\vert_{R_l}$.
The norm-conserving objective function is then formed as
\begin{equation}
    \mathcal{N}^2 = \sum_l (N^{\rm ECP}_l-N^{\rm AE}_l)^2 + (V^{\rm ECP}_l-V^{\rm AE}_l)^2 + (S^{\rm ECP}_l-S^{\rm AE}_l)^2 + (\epsilon^{\rm ECP}_l-\epsilon^{\rm AE}_l)^2,
\end{equation}
where we also include the errors in the orbital eigenvalues $\epsilon_l$.
This procedure is similar to what is generally done in norm-conserving schemes, however, we allow the core more freedom by not matching it to a particular functional form but rather focus on matching the quantities mentioned.
For the elements Al-Ar we took reference $s$ and $p$ orbitals from the atom's neutral ground state and took the reference $d$ orbital from the $p\rightarrow d$ excitation. 
For Na and Mg, we took the single-valence doublet states as the reference in each channel.

}

\subsection{Weighted Combination and Core-Valence Partitioning}

To form our full objective function, $O^2$, we take $\mathcal{E}^2$ and $\mathcal{N}^2$ as a linear combination,
\begin{equation}
O^2 = w_0\mathcal{E}^2+\mathcal{N}^2 ,
\end{equation}
{where $w_0$ is a chosen weight, typically taken as $5\times10^{-2}$}.
To construct our ECP, we vary its parameters until $O^2$ is minimized.
The value of $w_0$ is chosen so as to strike a balance between the accuracy of the many-body and single-body properties of the ECP -- or viewed in another way, we are allowing a violation of single-body norm conservation in order to achieve higher accuracies on the many-body spectral properties.

When using the conventional Ne-core and $3s3p$ valence, the transferability of the ECPs in particular environments was observed to be less than satisfying.
This is particularly true for compressed polar bonds (as will be shown).
In order to reach high many-body and single-body accuracies in these regimes as well, we have also generated He-core ECPs for all of the second row atoms.
Though, more expensive than their Ne-core counterparts the He-core ECPs still provide a non-negligible computational savings when considering their use in quantum Monte Carlo which scales with atomic number as $\mathcal{O}(Z^{5.5-6.5})$ and within planewave codes where calculations with the full Coulomb potential are not feasible.
{For these small-core ECPs, the optimization procedure was nearly identical, except that we took the neutral ground states for all references in the norm-conserving component and we generated the cutoff radii from the semi-core orbitals.
In this case, for each angular momentum channel, the cutoff radius was taken as the inner most finite extremum of {$r\phi^{\rm AE}_l$}.
We then used these radii as the match points of the norm, value and slope of the valence orbitals while also matching their eigenvalues. 
Furthermore, we found that a much larger weight could be used on the UCCSD(T) spectral component (two or three orders of magnitude) than the weight used for the Ne-core ECPs without introducing a significant negative impact in transferability.
This is due to a much smaller discrepancy in the correlation energies (valence and semi-core electrons) between the He-core and all-electron cases. 
} 

We have applied the construction outlined in this section to all the second row  atoms, Na-Ar.
For all optimizations, we have used the DONLP2 solver of Spellucci \cite{DONLP2} to generate the optimal parameter sets for all ECPs.
The methodologies used in DONLP2 are outlined in \cite{SpellucciA} and \cite{SpellucciB}.
Though finding the global minimum in these types of non-linear optimizations can be difficult, our incorporation of the many-body spectral component and mean-field spatial component (ie, additional constraints) should alleviate issues related to the ill-conditioned nature of this problem. 
Further justification of our chosen objective function came after analyzing the accuracy of our ECPs in settings that were not included as part of their training, which we cover in section VI.

The parameter values of our ccECPs for all atoms and all core sizes are shared in Tables I and II and are also available at the website in Ref. \cite{website}.

\begin{table*}[ht!]%
    \centering
    \scriptsize
    \caption{Parameter values for Ne-core ECPs. For all ECPs, the highest $l$ value corresponds to the local channel. 
    }
    \begin{minipage}{0.5\linewidth}

    \begin{tabular}{ccccrr}
	 \hline
	 \hline
     Atom &   $Z_{\mathrm{eff}}$ &  $l$ & $n_{l,k}$ &  \multicolumn{1}{c}{$\alpha_{l,k}$} & \multicolumn{1}{c}{$\beta_{l,k}$} \\
     \hline 
       Na &           1 &   0  &  2  & 5.377666  &  6.234064  \\
          &             &   0  &  2  & 1.408414  &  9.075931  \\
          &             &   1  &  2  & 1.379949  &  3.232724  \\
          &             &   1  &  2  & 0.862453  &  2.494079  \\
          &             &   2  &  1  & 4.311678  &  1.000000  \\
          &             &   2  &  3  & 1.925689  &  4.311678  \\
          &             &   2  &  2  & 1.549498  & -2.083137  \\
          &             &      &     &             &              \\    
       Mg &           2 &   0  &  2  & 5.936017 &   6.428631  \\
          &             &   0  &  2  & 1.592891 &  14.195491  \\
          &             &   1  &  2  & 1.583969 &   3.315069  \\
          &             &   1  &  2  & 1.077297 &   4.403025  \\
          &             &   2  &  1  & 6.048538 &   2.000000  \\
          &             &   2  &  3  & 2.796989 &  12.097075  \\
          &             &   2  &  2  & 2.547408 & -17.108313  \\
          &             &      &     &             &              \\    
       Al &           3 &   0  &  2  & {7.863954}  & { 14.879513}  \\
          &             &   0  &  2  & {2.061358}  & { 20.746863}  \\
          &             &   1  &  2  & {3.125175}  & {  7.786227}  \\
          &             &   1  &  2  & {1.414930}  & {  7.109015}  \\
          &             &   2  &  1  & {5.073893}  & {  3.000000}  \\
          &             &   2  &  3  & {8.607001}  & { 15.221680}  \\
          &             &   2  &  2  & {3.027490}  & {-11.165685}  \\
          &             &      &     &                &                 \\    
       Si &           4 &   0  &  2  & {9.447023}  & { 14.832760}  \\
          &             &   0  &  2  & {2.553812}  & { 26.349664}  \\
          &             &   1  &  2  & {3.660001}  & {  7.621400}  \\
          &             &   1  &  2  & {1.903653}  & { 10.331583}  \\
          &             &   2  &  1  & {5.168316}  & {  4.000000}  \\
          &             &   2  &  3  & {8.861690}  & { 20.673264}  \\
          &             &   2  &  2  & {3.933474}  & {-14.818174}  \\
	 \hline
	 \hline
    \end{tabular}\label{tab:ne_core_params}
    \end{minipage}%
    \begin{minipage}{0.5\linewidth}
    \begin{tabular}{ccccrr}
	 \hline
	 \hline
     Atom &   $Z_{\mathrm{eff}}$ &  $l$ & $n_{l,k}$ &  \multicolumn{1}{c}{ $\alpha_{l,k}$ } & \multicolumn{1}{c}{$\beta_{l,k}$} \\
     \hline
        P &           5 &   0  &  2  & {12.091334}  & { 15.259383}  \\
          &             &   0  &  2  & { 3.044535}  & { 31.707918}  \\
          &             &   1  &  2  & { 4.310884}  & {  7.747190}  \\
          &             &   1  &  2  & { 2.426903}  & { 13.932528}  \\
          &             &   2  &  1  & { 5.872694}  & {  5.000000}  \\
          &             &   2  &  3  & { 9.891298}  & { 29.363469}  \\
          &             &   2  &  2  & { 4.692469}  & {-17.011136}  \\
	  &             &      &     &             &              \\    
        S &           6 &   0  &  2  & {16.117687}  & { 15.925748}  \\
          &             &   0  &  2  & { 3.608629}  & { 38.515895}  \\
          &             &   1  &  2  & { 6.228956}  & {  8.062221}  \\
          &             &   1  &  2  & { 2.978074}  & { 18.737525}  \\
          &             &   2  &  1  & { 6.151144}  & {  6.000000}  \\
          &             &   2  &  3  & {11.561575}  & { 36.906864}  \\
          &             &   2  &  2  & { 5.390961}  & {-19.819533}  \\
          &             &      &     &             &              \\    
       Cl &           7 &   0  &  2  & {17.908432}  & { 15.839234}  \\
          &             &   0  &  2  & { 4.159880}  & { 44.469504}  \\
          &             &   1  &  2  & { 7.931763}  & {  8.321946}  \\
          &             &   1  &  2  & { 3.610412}  & { 24.044745}  \\
          &             &   2  &  1  & { 7.944352}  & {  7.000000}  \\
          &             &   2  &  3  & {12.801261}  & { 55.610463}  \\
          &             &   2  &  2  & { 6.296744}  & {-22.860784}  \\
          &             &      &     &             &              \\    
       Ar &           8 &   0  &  2  &  {27.068139}  & { 18.910152 }  \\
          &             &   0  &  2  &  { 4.801263}  & { 53.040012 }  \\
          &             &   1  &  2  &  {11.135735}  & {  8.015534 }  \\
          &             &   1  &  2  &  { 4.126631}  & { 28.220208 }  \\
          &             &   2  &  1  &  { 8.317181}  & {  8.000000 }  \\
          &             &   2  &  3  &  {13.124648}  & { 66.537451 }  \\
          &             &   2  &  2  &  { 6.503132}  & {-24.100393 }  \\   
	 \hline
	 \hline
    \end{tabular}\label{tab:he_core_params}
    \end{minipage}
    \label{tab:ecp_params}%
\end{table*}

\begin{table*}[ht!]%
    \centering
    \scriptsize
    \caption{Parameter values for He-core ECPs. For all ECPs, the highest $l$ value corresponds to the local channel. 
    }
    \begin{minipage}{0.5\linewidth}
    \begin{tabular}{ccccrr}
	 \hline
	 \hline
     Atom &   $Z_{\mathrm{eff}}$ &  $l$ & $n_{l,k}$ &  \multicolumn{1}{c}{$\alpha_{l,k}$} & \multicolumn{1}{c}{ $\beta_{l,k}$ } \\
     \hline  
       Na &           9 &   0  &  2  &  {61.004364}   & { 25.194598} \\
          &             &   0  &  2  &  {14.829519}   & { 60.933576} \\
          &             &   1  &  1  &  { 8.997115}   & {  9.000000} \\
          &             &   1  &  3  &  { 8.958240}   & { 80.974036} \\
          &             &   1  &  2  &  { 8.691287}   & {-57.117991} \\
          &             &      &     &               &               \\    
       Mg &          10 &   0  &  2  &  {72.718763}   & { 25.152097} \\
          &             &   0  &  2  &  {19.375648}   & { 84.969927} \\
          &             &   1  &  1  &  {10.136825}   & { 10.000000} \\
          &             &   1  &  3  &  {10.457653}   & {101.368253} \\   
          &             &   1  &  2  &  {10.109595}   & {-68.596401} \\
          &             &      &     &               &               \\    
        Al &         11 &   0  &  2  &   {81.815564}   & { 25.157259} \\
          &             &   0  &  2  &   {24.522883}   & {113.067525} \\
          &             &   1  &  1  &   {11.062056}   & { 11.000000} \\
          &             &   1  &  3  &   {12.369778}   & {121.682619} \\  
          &             &   1  &  2  &   {11.965444}   & {-82.624567} \\  
          &             &      &     &               &               \\    
       Si &          12 &   0  &  2  &  {92.046246}   & { 25.228329} \\
          &             &   0  &  2  &  {30.895726}   & {150.483122} \\
          &             &   1  &  1  &  {12.066048}   & { 12.000000} \\
          &             &   1  &  3  &  {15.276621}   & {144.792575} \\
          &             &   1  &  2  &  {14.506273}   & {-99.229393} \\ 
          &             &      &     &               &               \\   
	 \hline
	 \hline
    \end{tabular}\label{tab:ne_core_params}
    \end{minipage}%
    \begin{minipage}{0.5\linewidth}
    \begin{tabular}{ccccrr}
	 \hline
	 \hline
     Atom &   $Z_{\mathrm{eff}}$ &  $l$ & $n_{l,k}$ &  \multicolumn{1}{c}{ $\alpha_{l,k}$ } & \multicolumn{1}{c}{$\beta_{l,k}$} \\
     \hline  
        P &          13 &   0  &  2  &  {101.982019}   & {  25.197230} \\
          &             &   0  &  2  &  { 37.485881}   & { 189.426261} \\
          &             &   1  &  1  &  { 15.073300}   & {  13.000000} \\
          &             &   1  &  3  &  { 18.113176}   & { 195.952906} \\ 
          &             &   1  &  2  &  { 17.371539}   & {-117.611086} \\
          &             &      &     &               &               \\    
        S &          14 &   0  &  2  &  {111.936344}   & {  25.243283} \\
          &             &   0  &  2  &  { 43.941844}   & { 227.060768} \\
          &             &   1  &  1  &  { 17.977612}   & {  14.000000} \\
          &             &   1  &  3  &  { 20.435964}   & { 251.686565} \\
          &             &   1  &  2  &  { 19.796579}   & {-135.538891} \\
          &             &      &     &               &               \\    
       Cl &          15 &   0  &  2  &  {124.640433}   & {  26.837357} \\
          &             &   0  &  2  &  { 52.205433}   & { 277.296696} \\
          &             &   1  &  1  &  { 22.196266}   & {  15.000000} \\
          &             &   1  &  3  &  { 26.145117}   & { 332.943994} \\
          &             &   1  &  2  &  { 25.015118}   & {-161.999982} \\
          &             &      &     &               &               \\    
       Ar &          16 &   0  &  2  &  {135.620522}   & {  25.069215} \\
          &             &   0  &  2  &  { 60.471053}   & { 332.151842} \\
          &             &   1  &  1  &  { 23.431337}   & {  16.000000} \\
          &             &   1  &  3  &  { 26.735872}   & { 374.901386} \\
          &             &   1  &  2  &  { 26.003325}   & {-178.039517} \\
          &             &      &     &               &               \\           
	 \hline
	 \hline
    \end{tabular}\label{tab:he_core_params}
    \end{minipage}
    \label{tab:ecp_params}%
\end{table*}

\section{ECP Form}

The ECPs we have constructed have the following form
\begin{equation}
    V^{\rm ECP }_i = V_{\rm loc}(r_i) + \sum_{l=0}^{l_{\rm max}} V_l(r_i) \sum_{m}|lm\rangle\langle lm|,
\end{equation}
where $r_i$ is the radial distance of electron $i$ from the core's origin and $l_{\rm max}$ is chosen to be the maximum occupied angular momentum channel from the ground state core. 
The non-local terms include $lm$-th angular momentum projectors, $|lm\rangle\langle lm|$.

The local potential, $V_{\rm loc}$, is taken to be
\begin{equation}
\begin{aligned}
    V_{\rm loc}(r) &= -\frac{Z_{\rm eff}}{r}+\sum_{k=1} \beta_{{\rm loc},k}r^{n_{{\rm loc},k}-2}e^{-\alpha_{{\rm loc},k}r^2} \\
                   &= -\frac{Z_{\rm eff}}{r}+\frac{\beta_{{\rm loc},1}}{r}e^{-\alpha_{{\rm loc},1} r^2} + \beta_{{\rm loc},2}re^{-\alpha_{{\rm loc},2} r^2} + \sum_{k=3}\beta_{{\rm loc},k}e^{-\alpha_{{\rm loc},k} r^2},
\end{aligned}
\end{equation}
where $Z_{\rm eff}=Z-Z_{core}$ is the pseudo-atom's total core charge.
We fix $\beta_{{\rm loc},1}=Z_{\rm eff}$, $\beta_{{\rm loc},2}=Z_{\rm eff}\alpha_{{\rm loc},1}$,  while $n_{loc,1}=1, n_{loc,2}=3$, so that $V_{\rm loc}(r)$ is finite everywhere and its
first derivative vanishes at the origin.
Similarly, the set of non-local potentials, $\{V_l\}$, are parameterized by
\begin{equation}
    V_l(r) = \sum_{k=1} \beta_{l,k}r^{n_{l,k}-2}e^{-\alpha_{l,k}r^2},
\end{equation}
where the powers $\{n_{l,k}\}$, are all equal to $2$.
Except for $\beta_{loc,1}$ and
$\beta_{loc,2}$, which are fixed as given above, the Greek letter parameters are all varied in the minimization of our objective function.
For an initial round of optimization, for $V_{\rm loc}$ and all $V_l$'s, we set the number of gaussian terms to one.
For a given Ne-core ECP, we included 1-2 more gaussians in these channels for additional flexibility if further optimization rounds were required to reach agreement within roughly 0.05-0.1 eV of the low-lying all-electron many-body spectrum.
Pushing the accuracies further, generally led to significant mean-field errors for reasons discussed in section II.
In the case of the He-core ECPs, we again only included one gaussian term in these functions initially. However, for these ECPs further terms were added until we observed agreement with all-electron data within roughly the chemical accuracy window.
This simple form keeps the volume of the parameter space as small as practically possible and therefore helps to simplify the optimization problem.

\section{Valence basis sets}
For each atom and each core size, we have generated a corresponding basis set.
The general recipe that we followed to generate the basis sets (similar to Ref. \cite{Burkatzki:2007jcp}) was the same for both core sizes and we outline it in this section.

To generate the basis sets for the He-core ECPs, we have minimized the atomic ground-state CCSD(T) energy for all atoms.
The basis sets for the Ne-core ECPs minimize the atomic ground-state CCSD(T) energy for the atoms Al through Ar and the dimer ground-state CCSD(T) energy for the atoms Na and Mg. 
For a given basis set, we start by generating contractions, one for each occupied HF ground-state orbital, by starting from a set of {primitives and optimizing the exponents until the ground-state HF energy is minimized.}
The resulting exponents are then taken as the contraction primitives whose coefficients are taken to be the ground-state HF expansion coefficients.
We take the same contractions for each VDZ, VTZ, VQZ and V5Z basis.
For a given basis, we append additional even-tempered uncontracted primitives following the correlation-consistent scheme \cite{dunningcc} and determine the largest exponent among these primitives by minimizing the ground-state CCSD(T) energy.

In addition to these basis sets, we have also generated augmented varieties which follow the same scheme, but include an additional even-tempered diffuse primitive into each symmetry channel.
All of our basis sets are shared at the website given in Ref. \cite{website}.

\section{Results}
For all 2nd row atoms, we share in Tables \ref{tab:na_atomic}, \ref{tab:mg_atomic}, \ref{tab:al_atomic}, \ref{tab:si_atomic}, \ref{tab:p_atomic}, \ref{tab:s_atomic},\ref{tab:cl_atomic} and \ref{tab:ar_atomic} the all-electron CCSD(T) valence ionization potentials and (if bound) the electron affinity along with the discrepancies of the same quantities from various core approximations which includes all-electron UC approximation, a number of previously tabulated ECPs \cite{SBK,Burkatzki:2007jcp,STU,Trail:2005jcp} and our Ne-core and He-core correlation consistent ECPs, ccECP[He] and ccECP[Ne], respectively.
We also share the discrepancies from all-electron UCCSD(T) potential energy surfaces for one or two diatomic molecules for all the atoms in Figs. \ref{fig:Na_mols}, \ref{fig:MgO}, \ref{fig:Al_mols}, \ref{fig:Si_mols}, \ref{fig:P_mols}, \ref{fig:S_mols}, \ref{fig:Cl_mols} and \ref{fig:ArH}.
For the cases of oxides, we utilized our previously published oxygen ECP \cite{Bennett:2017jcp} when calculating potential energy surfaces for our ECPs and for all other ECPs we used that table's corresponding oxygen ECP.
For the molecular discrepancies, we span a large range of geometries in all cases, from near the dissociation threshold ($\lessapprox 0.05$ \AA) on the left to just past equilibrium.
For argonium, we used a well-tuned unpublished hydrogen ECP for all cases other than uncorrelated core. 
Our ECP parameter sets are shared in Tables \ref{tab:ne_core_params} and \ref{tab:he_core_params}.

In the following subsections, we provide a discussion for each pseudoatom's atomic and molecular accuracies, in turn, and summarize the results.
In the final subsection, we share a comparison of the mean absolute deviations of the binding parameters of the molecules presented in this work between our ECPs and the all-electron UC approximation as well as the other tabulated ECPs we've benchmarked against in this work to provide a more global picture of the accuracy of our potentials. In this section we also present total atomic energy components of our ECPs along with their core radii for future reference.

\subsection{Sodium}

\begin{figure*}[h!]
\centering
\begin{subfigure}{0.5\textwidth}
\includegraphics[width=\textwidth]{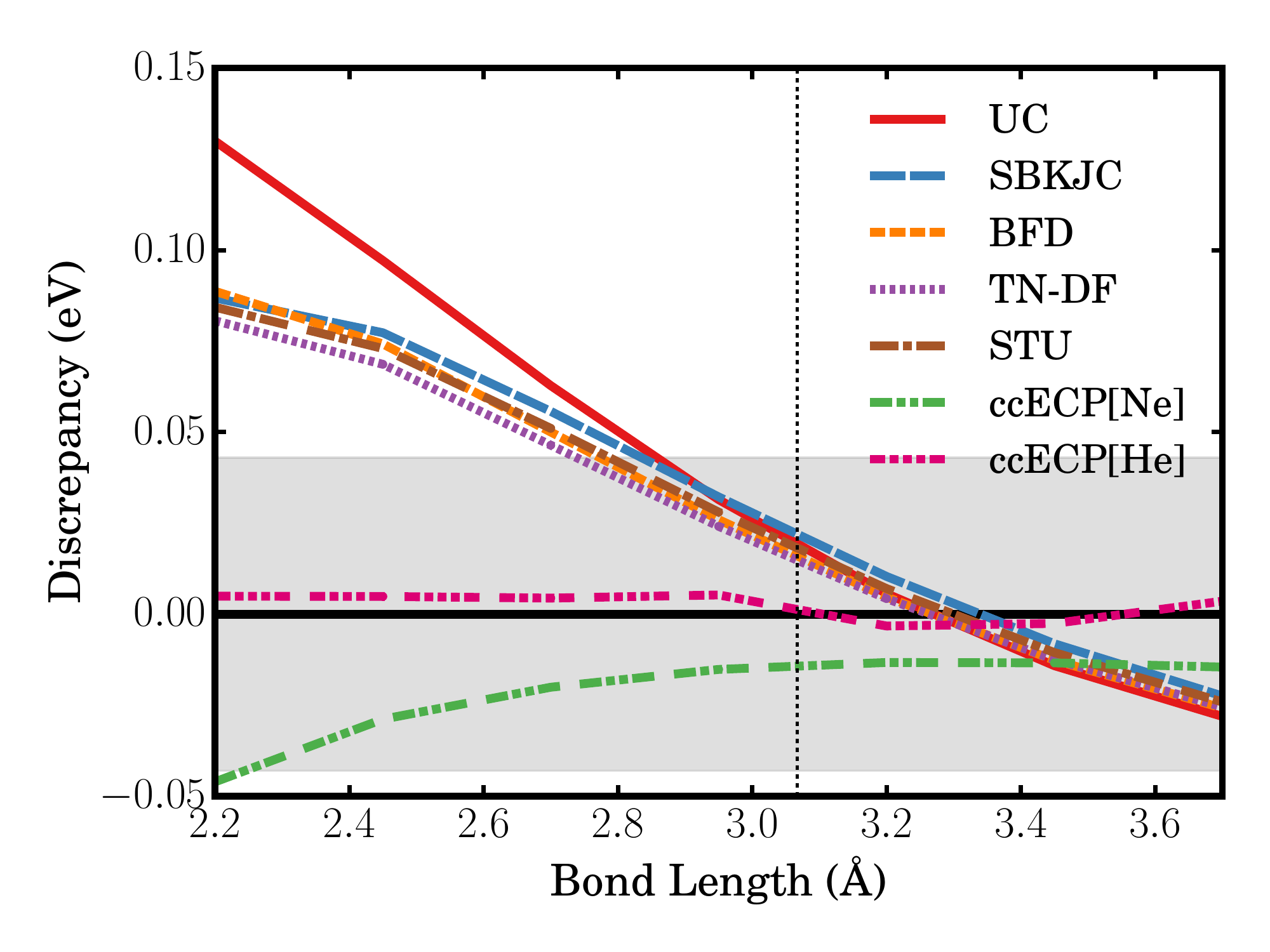}
\caption{Na$_2$ binding curve discrepancies}
\label{fig:Na2}
\end{subfigure}%
\begin{subfigure}{0.5\textwidth}
\includegraphics[width=\textwidth]{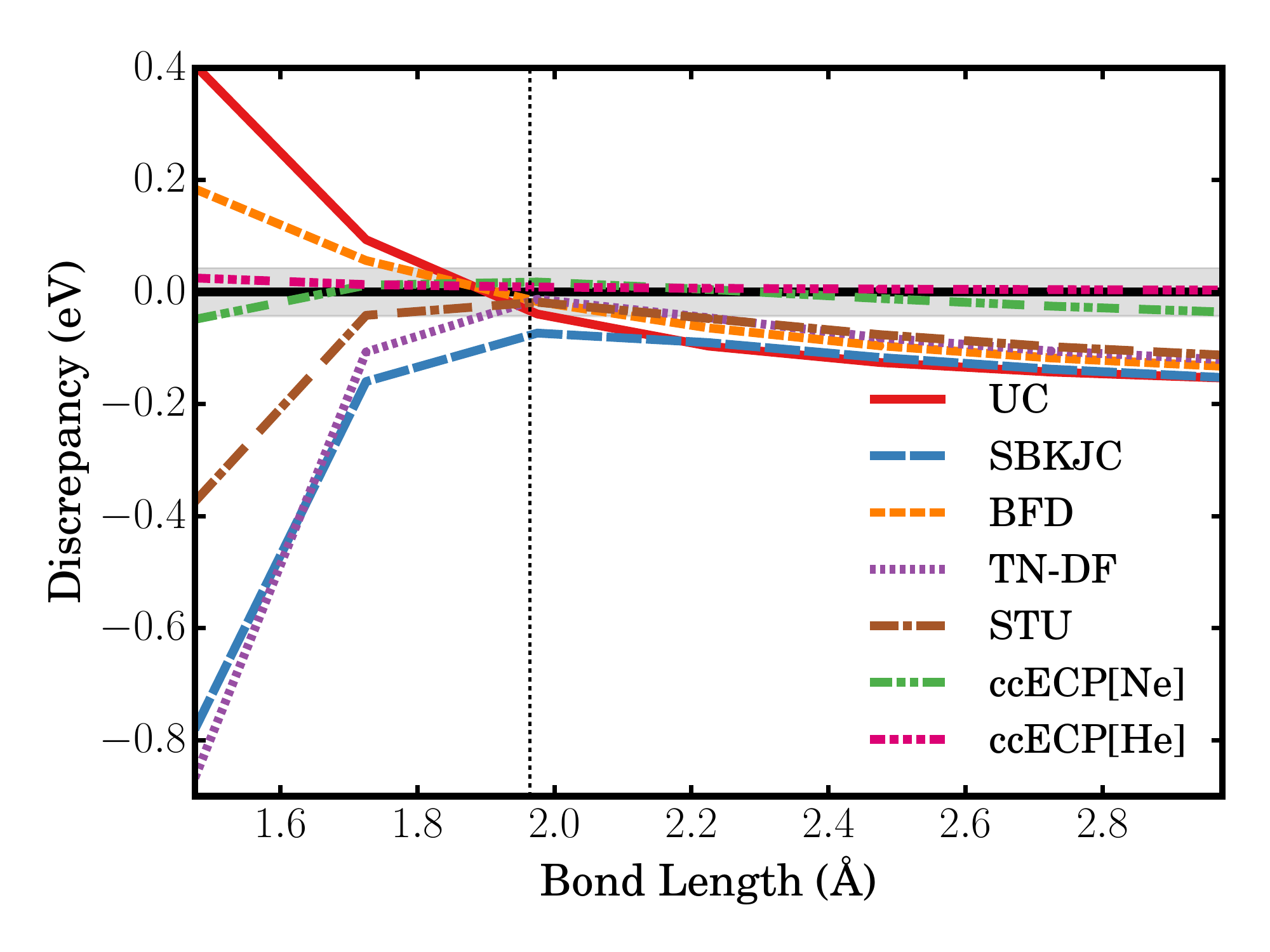}
\caption{NaO binding curve discrepancies}
\label{fig:NaO}
\end{subfigure}
\caption{Binding energy discrepancies for (a) Na$_2$ and (b) NaO molecules in their ground states $^1\Sigma_g$ and $^2\Sigma$, respectively. The binding curves are relative to the AE UCCSD(T) binding curve. The shaded region indicates a discrepancy of chemical accuracy in either direction. }
\label{fig:Na_mols}
\end{figure*}

In the case of sodium, the atomic and molecular results of our Ne-core and He-core ECPs are shared in Table \ref{tab:na_atomic} and Fig.\ref{fig:Na_mols}, respectively. 
For the case of the Ne-core, our construction led to a significant improvement of the atom's first ionization potential, when compared to the all-electron UCCSD(T) result, relative to the previously tabulated ECPs as well as the uncorrelated core (UC) result -- the error being roughly a factor of $3$ smaller than the other ECPs and roughly a factor of $2$ smaller than UC.
The electron affinity of our Ne-core ECP is also in good agreement with the all-electron result with an error that is less than $0.01$ eV.
In the case of our He-core ECP, the agreement with the all-electron atom is better still and for each quantity, the errors are at sub meV scales.

For the Na$_2$ and NaO molecules, the accuracies of both of our core partitions are also quite high with our Ne-core and He-core ECPs being within $0.05$ eV and $0.01$ eV of the all-electron UCCSD(T) result over the full range of geometries, respectively.

\begin{table*}[h!]
    \centering
    \caption{All-electron (AE) UCCSD(T) electron affinity and ionization potential of Na along with the errors from uncorrelated core (UC), ECPs and for information purposes also from experiment (Exp.). The uncontracted aug-cc-pCV5Z basis was used for all calculations. MAD is the mean absolute deviation of excitation energies, while MARE is the mean absolute relative error. All values in eV.}
    \label{tab:na_atomic}
    \setlength{\tabcolsep}{5pt}
    \begin{tabular}{lrrrrrrrrrrr}
         \hline
         \hline
           \multicolumn{1}{l}{\multirow{2}{*}{Qty.}} & \multicolumn{1}{r}{\multirow{2}{*}{Exp.}} & \multicolumn{1}{r}{\multirow{2}{*}{AE}} &  \multicolumn{7}{c}{Discrepancies from AE}  \\
         &  &  & \multicolumn{1}{r}{UC} & \multicolumn{1}{r}{SBKJC} &     \multicolumn{1}{r}{BFD} &   \multicolumn{1}{r}{TN-DF} &     \multicolumn{1}{r}{STU} & \multicolumn{1}{r}{ccECP[Ne]} & \multicolumn{1}{r}{ccECP[He]} \\
     \hline
          IP(I) & 5.1391$^a$     & 5.1334 &   -0.1215 & -0.1847 & -0.1770 & -0.1738 & -0.1774 & -0.0665 & {-0.0020} \\
             EA & 0.5479$^b$     & 0.5470 &   -0.0017 &  0.0005 &  0.0030 &  0.0025 &  0.0026 &  0.0077 & {-0.0014} \\
         \hline
        \multicolumn{3}{l}{ MAD }         &     0.0616 &     0.0926 &     0.0900 &     0.0882 &     0.0900 &     0.0371 &    {0.0017} \\ 
        \multicolumn{3}{l}{ MARE }       &     0.0134 &     0.0184 &     0.0200 &     0.0192 &     0.0197 &     0.0135 &     {0.0015} \\
         \hline
         \hline
         \multicolumn{10}{l}{\footnotesize $^a$ Reference \cite{NIST_ASD}} \\ 
         \multicolumn{10}{l}{\footnotesize $^b$ Reference \cite{EA}}
    \end{tabular}
\end{table*}

\subsection{Magnesium}
\begin{figure}[h!]
\centering
\caption{Binding energy discrepancies for the MgO molecule in its ground state $^1\Sigma^{+}$. The binding curves are relative to the AE UCCSD(T) binding curve. The shaded region indicates a discrepancy of chemical accuracy in either direction. }
\includegraphics[width=0.5\textwidth]{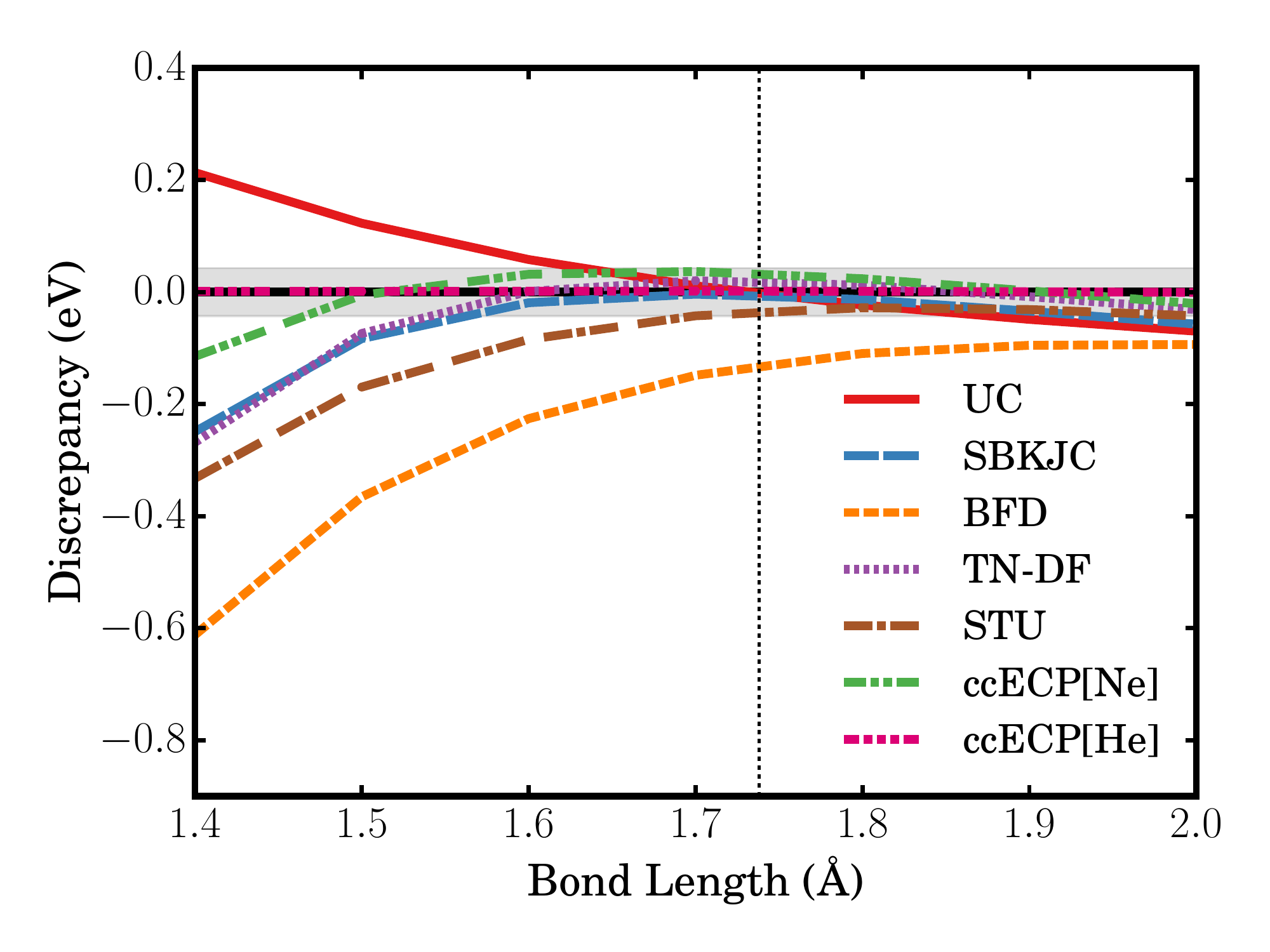}
\label{fig:MgO}
\end{figure}

For magnesium, the atomic and molecular results of our Ne-core and He-core ECPs are shared in Table \ref{tab:mg_atomic} and Fig. \ref{fig:MgO}, respectively.
The Ne-core partition shows an increased accuracy over UC and the previously tabulated ECPs with respect to the first and second ionization potentials.
The He-core partition further increases the accuracy by more than an order of magnitude, with errors less than a meV for these quantities.

In the case of the MgO molecule, our Ne-core ECP shows better agreement with all-electron UCCSD(T) at shorter atomic separations than both UC and previously tabulated ECPs and at equilibrium, the agreement is within chemical accuracy.
Our He-core ECP improves upon this further, with errors less than $0.01$ eV across all geometries plotted.

\begin{table*}[h!]
    \centering
    \caption{All-electron UCCSD(T) ionization potentials for Mg along with the errors from uncorrelated core (UC) and ECPs. The uncontracted aug-cc-pCV5Z basis was used for all calculations. All values in eV. See Tab. III for further description.}
    \begin{tabular}{lrrrrrrrrrrrr}
           \hline
           \hline
           \multicolumn{1}{l}{\multirow{2}{*}{Qty.}} & \multicolumn{1}{r}{\multirow{2}{*}{Exp.}} & \multicolumn{1}{r}{\multirow{2}{*}{AE}} &  \multicolumn{7}{c}{Discrepancies from AE}  \\
         &  &  & \multicolumn{1}{r}{UC} & \multicolumn{1}{r}{SBKJC} &     \multicolumn{1}{r}{BFD} &   \multicolumn{1}{r}{TN-DF} &     \multicolumn{1}{r}{STU} & \multicolumn{1}{r}{ccECP[Ne]} & \multicolumn{1}{r}{ccECP[He]} \\
           \hline
            IP(I) &  7.6462$^a$  &  7.6400 & -0.0980 & -0.0884 & -0.0727 & -0.0693 & -0.0617 &    -0.0578 &    { 0.0059} \\
           IP(II) & 15.0353$^a$  & 15.0287 & -0.2858 & -0.3003 & -0.2898 & -0.2709 & -0.2757 &    -0.2050 &    {-0.0050} \\
         \hline
        \multicolumn{3}{l}{ MAD }         &     0.1919 &     0.1944 &     0.1812 &     0.1701 &     0.1687 &     0.1314 &  {0.0055} \\ 
        \multicolumn{3}{l}{ MARE }         &     0.0159 &     0.0158 &     0.0144 &     0.0135 &     0.0132 &     0.0106 &     {0.0006} \\ 
         \hline
         \hline
          \multicolumn{10}{l}{\footnotesize $^a$ Reference \cite{NIST_ASD}} 
    \end{tabular}
    \label{tab:mg_atomic}
\end{table*}

\subsection{Aluminum}
\begin{figure*}[!h]
\centering
\begin{subfigure}{0.5\textwidth}
\includegraphics[width=\textwidth]{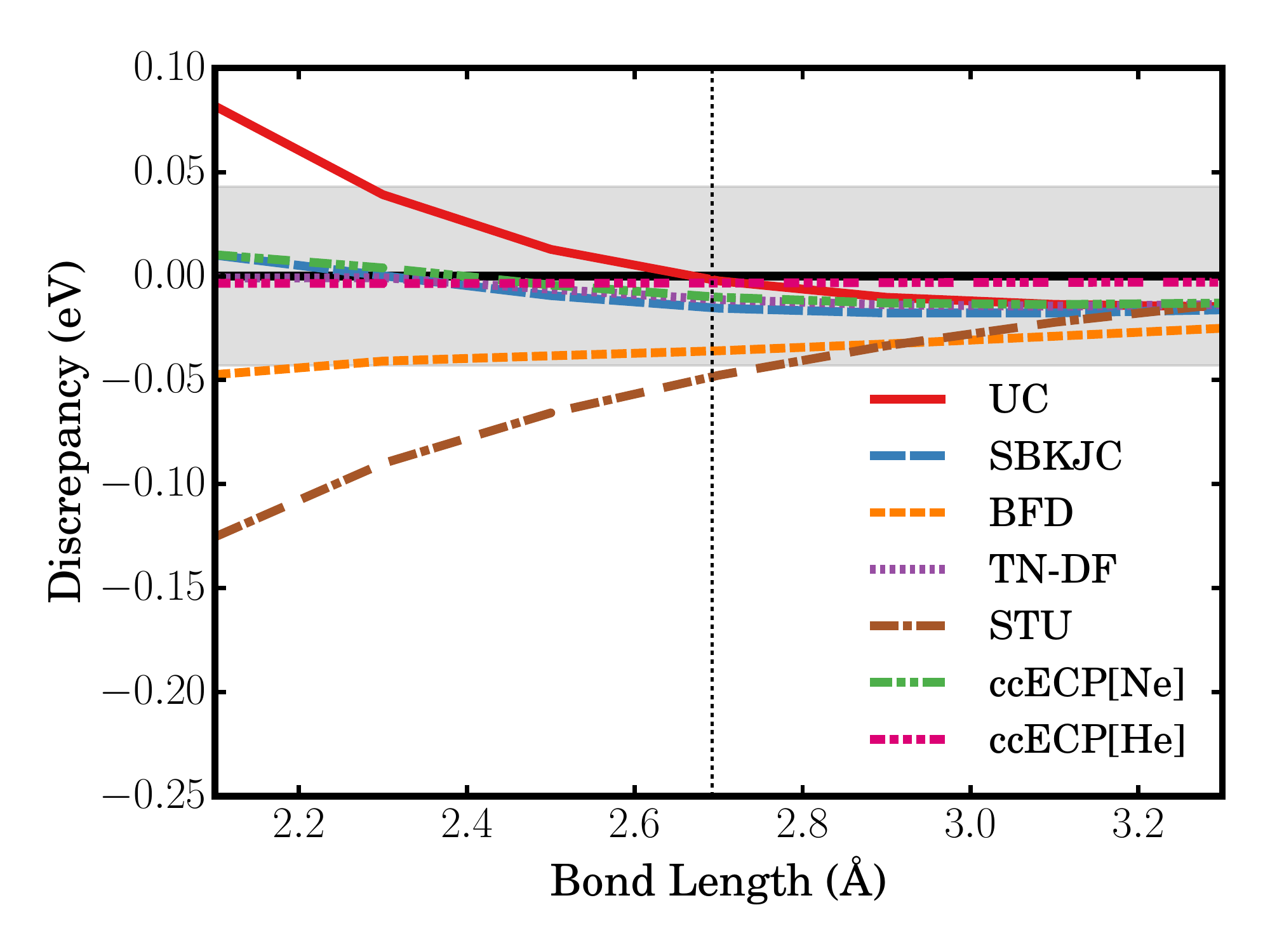}
\caption{Al2 binding curve discrepancies}
\label{fig:Al2}
\end{subfigure}%
\begin{subfigure}{0.5\textwidth}
\includegraphics[width=\textwidth]{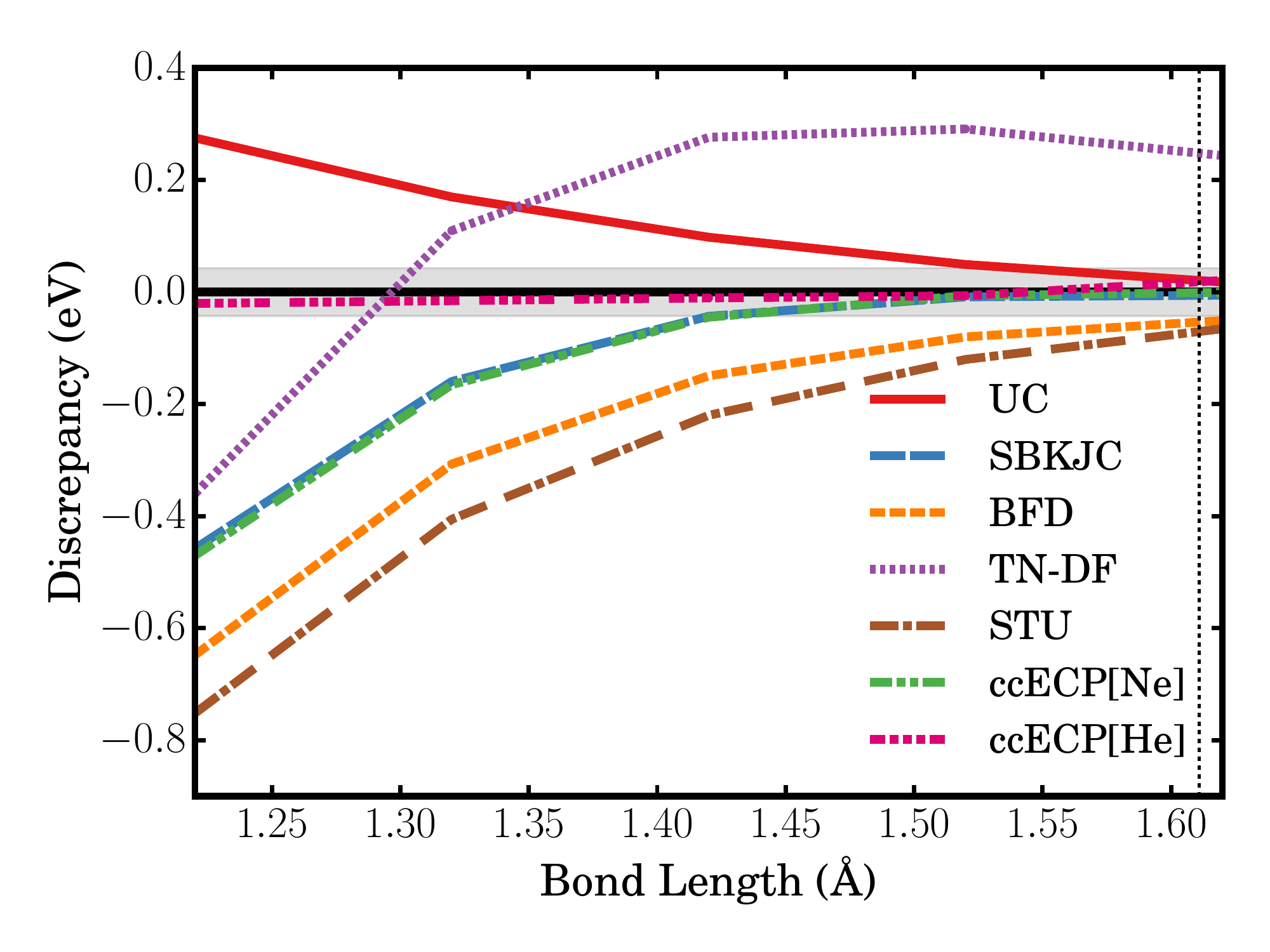}
\caption{AlO binding curve discrepancies}
\label{fig:AlO}
\end{subfigure}
\caption{Binding energy discrepancies for (a) Al2 and (b) AlO molecules in their ground states $^3\Sigma_g$ and $^2\Sigma$, respectively. The binding curves are relative to the AE UCCSD(T) binding curve. The shaded region indicates a discrepancy of chemical accuracy in either direction. }
\label{fig:Al_mols}
\end{figure*}

For aluminum, the atomic and molecular results of our Ne-core and He-core ECPs are shared in Table \ref{tab:al_atomic} and Fig. \ref{fig:Al_mols}, respectively.
In the Ne-core case, the errors over the ionization potentials and the electron affinity were not improved over the other core approximations but instead remained comparable. We note the
lower variational freedom for our case due to bounded forms and smaller radial range when compared with some of the constructions.
However, in the case of the He-core, we observed errors less than a hundredth of an eV.

The Al$_2$ molecule is described well by both core partitions, both being within chemical accuracy scales.
For the AlO molecule, similar to previously tabulated ECPs, our Ne-core ECP is overbound by tenths of eV as the molecule is compressed towards dissociation.
This potentially signifies a limitation of this particular choice of core for this type of environment, namely, compressed polar bonds.
We suspect that key contributors to the error in this regime are a breakdown of the assumed point-charge interactions of the cores and also the neglect core-valence overlaps \cite{DolgCao,Chang:1977tca}.
For our smaller He-core ECP, we see that these difficulties evaporate and we observe errors well within chemical accuracy. 
\begin{table*}[h!]
    \centering
    \caption{All-electron (AE) UCCSD(T) ionization potentials and electron affinity of Al along with the errors from the uncorrelated core (UC) and ECPs. Exp. gives experimental values for information purposes. The uncontracted aug-cc-pCV5Z basis was used for all calculations. AMAD is the mean absolute deviation for all excitation energies, LMAD is the mean absolute deviation for the first and second ionization potentials and electron affinity, while MARE is the mean absolute relative error for all states. All values in eV.}
    \begin{tabular}{lrrrrrrrrrrrr}
           \hline
           \hline
           \multicolumn{1}{l}{\multirow{2}{*}{Qty.}} & \multicolumn{1}{r}{\multirow{2}{*}{Exp.}} &  \multicolumn{1}{r}{\multirow{2}{*}{AE}} &  \multicolumn{7}{c}{Discrepancies from AE}  \\
         &  &  & \multicolumn{1}{r}{UC} & \multicolumn{1}{r}{SBKJC} &     \multicolumn{1}{r}{BFD} &   \multicolumn{1}{r}{TN-DF} &     \multicolumn{1}{r}{STU} & \multicolumn{1}{r}{ccECP[Ne]} & \multicolumn{1}{r}{ccECP[He]} \\
           \hline
               IP(I)  &  5.9858$^a$   &  5.9606 &  0.0045 &  0.0214 &  0.0135 &  0.0167 &  0.0667 & { 0.0157} &  {-0.0027}  \\
              IP(II) &  18.8286$^a$   & 18.8216 & -0.1698 & -0.1720 & -0.1588 & -0.1408 & -0.0862 & {-0.1491} &  {-0.0051}  \\
             IP(III)  &  28.4476$^a$  & 28.4447 & -0.3588 & -0.4331 & -0.3339 & -0.3640 & -0.3384 & {-0.3915} &  { 0.0052}  \\
                  EA  &  0.4328$^b$   &  0.4183 & -0.0165 &  0.0215 &  0.0225 &  0.0182 &  0.0294 & { 0.0171} &  { 0.0010}  \\
         \hline
\multicolumn{3}{l}{ AMAD }               &     0.1374 &     0.1620 &     0.1322 &     0.1349 &     0.1302 &     { 0.1434} &     { 0.0035} \\ 
\multicolumn{3}{l}{ LMAD }               &     0.0636  &0.0716  &0.0649  &0.0586  &0.0608  & { 0.0606}  & { 0.0029}            \\
\multicolumn{3}{l}{ MARE }       &     0.0155 &     0.0198 &     0.0191 &     0.0166 &     0.0245 &     { 0.0163} &  { 0.0008} \\ 
         \hline
         \hline
          \multicolumn{10}{l}{\footnotesize $^a$ Reference \cite{NIST_ASD}} \\ 
         \multicolumn{10}{l}{\footnotesize $^b$ Reference \cite{EA}}
    \end{tabular}
    \label{tab:al_atomic}
\end{table*}

\subsection{Silicon}
\begin{figure*}[!h]
\centering
\begin{subfigure}{0.5\textwidth}
\includegraphics[width=\textwidth]{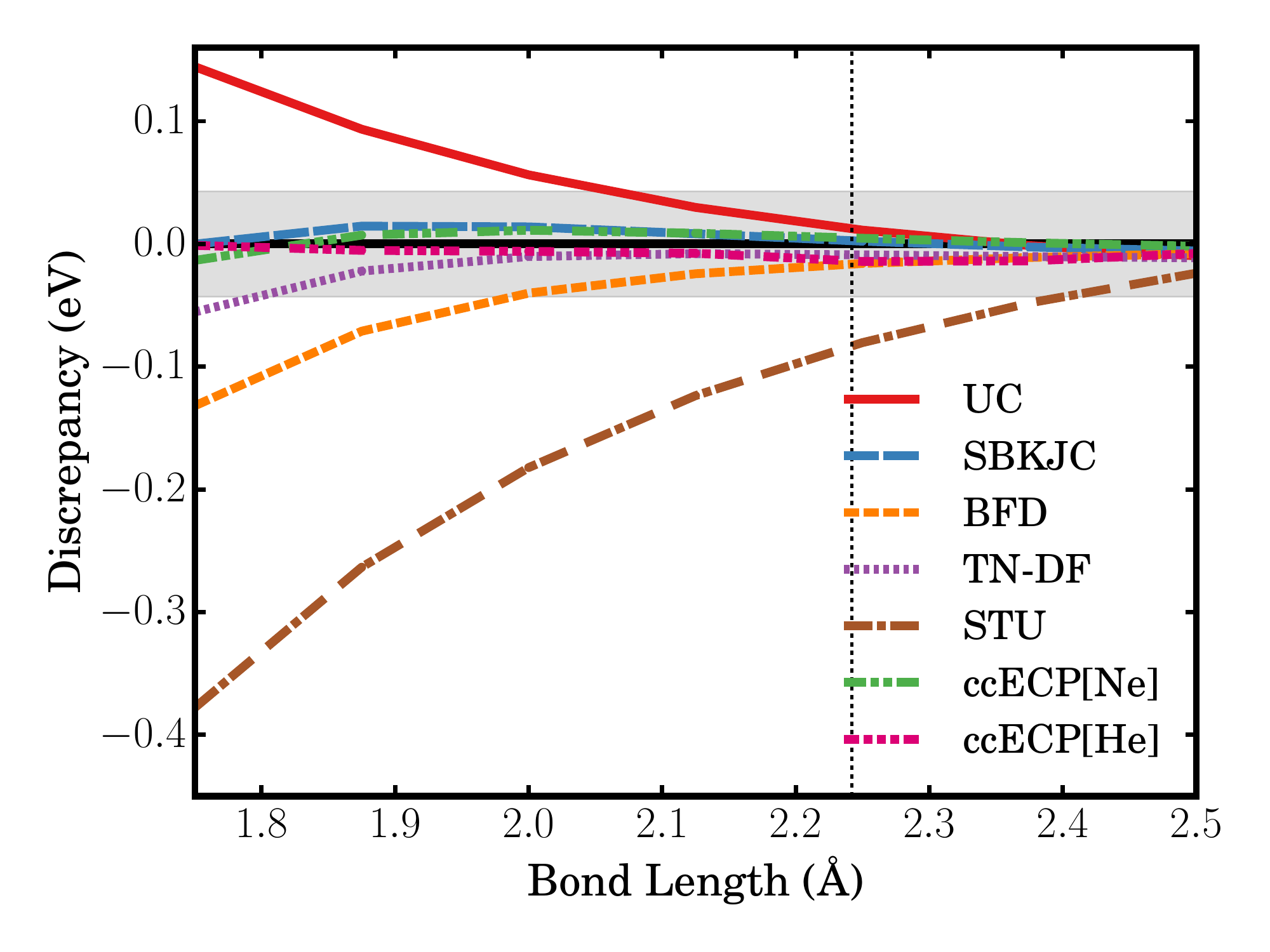}
\caption{Si2 binding curve discrepancies}
\label{fig:Si2}
\end{subfigure}%
\begin{subfigure}{0.5\textwidth}
\includegraphics[width=\textwidth]{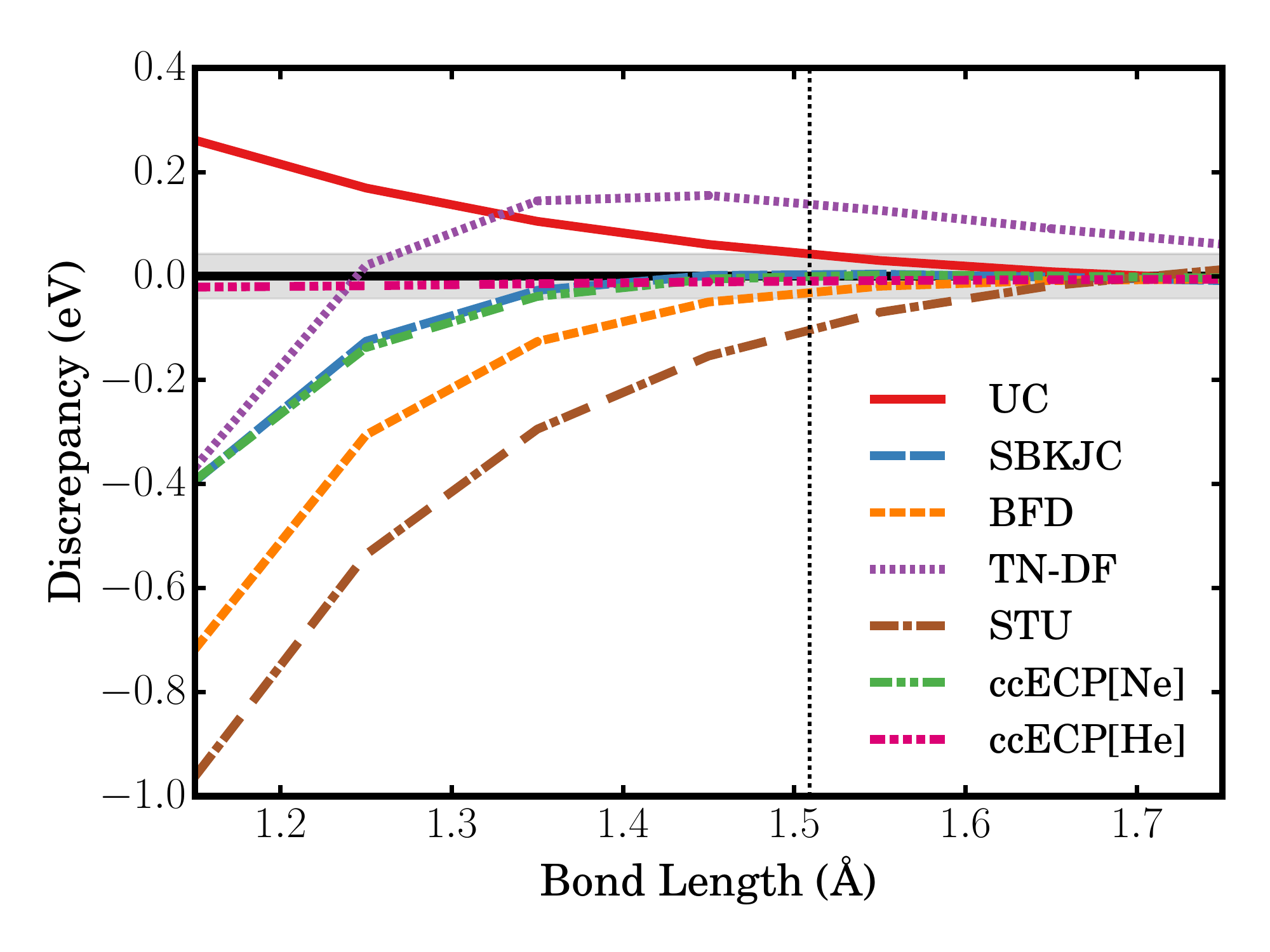}
\caption{SiO binding curve discrepancies}
\label{fig:SiO}
\end{subfigure}
\caption{Binding energy discrepancies for (a) Si2 and (b) SiO molecules in their ground states $^3\Sigma_g$ and $^1\Sigma$, respectively. The binding curves are relative to the AE UCCSD(T) binding curve. The shaded region indicates a discrepancy of chemical accuracy in either direction. }
\label{fig:Si_mols}
\end{figure*}

For silicon, the atomic and molecular results of our Ne-core and He-core ECPs are shared in Table \ref{tab:si_atomic} and Fig. \ref{fig:Si_mols}, respectively.
In the Ne-core case, like Al, the accuracies among the ionization potentials and electron affinity were not improved over the other core approximations, though the He-core ECP shows errors smaller than chemical accuracy for all of these quantities. The results from the Ne-core ECP potentially reveal the limitations of this particular core partition.

In the case of the silicon dimer, both core partitions perform very well and their errors at the UCCSD(T) level are no more than about $0.02$ eV.
The SiO molecule shows similar and profound discrepancies for the Ne-core ECP as was seen in AlO.
At short atomic separations, the pseudo-molecule is severely overbound for all tested ECPs including our Ne-core partition.
Interestingly, at equilibrium, the agreement for our ECP and some others is much better and is within chemical accuracy. 
For our He-core ECP, the errors are significantly reduced and we observe chemical accuracy agreement across all geometries plotted.  
On this element, it is clearly visible that 
atomic spectrum is not sufficient to assess
the overall quality of the ECPs. Seemingly,
BFD, TN, and STU have somewhat smaller
energy discrepancies than our Ne-core ECP.
However, molecular calculations reveal
lower or even significantly lower transferability. In fact, our Ne-core ECP is on 
par with SBKJC that however is unbounded 
($1/r^2$ terms) so that our variational freedom is then significantly lower in comparison.

\begin{table*}[h!]
    \centering
    \caption{All-electron (AE) UCCSD(T) ionization potentials and electron affinity of Si along with the errors from uncorrelated core (UC) and ECPs. The uncontracted aug-cc-pCV5Z basis was used for all calculations. All values in eV. See Tab. V for further description.}
    \begin{tabular}{lrrrrrrrrrrr}
           \hline
           \hline
           \multicolumn{1}{l}{\multirow{2}{*}{Qty.}} & \multicolumn{1}{r}{\multirow{2}{*}{Exp.}} &   \multicolumn{1}{r}{\multirow{2}{*}{AE}} &  \multicolumn{7}{c}{Discrepancies from AE}  \\
         &  &  & \multicolumn{1}{r}{UC} & \multicolumn{1}{r}{SBKJC} &     \multicolumn{1}{r}{BFD} &   \multicolumn{1}{r}{TN-DF} &     \multicolumn{1}{r}{STU} & \multicolumn{1}{r}{ccECP[Ne]} & \multicolumn{1}{r}{ccECP[He]} \\
           \hline
            IP(I)  &  8.1517$^a$   &  8.1392 & -0.0053 &  0.0190 &  0.0152 &  0.0183 &  0.0977 & { 0.0166} &  {-0.0011}        \\
           IP(II)  &  16.3459$^a$  & 16.3014 & -0.0320 & -0.0096 & -0.0131 & -0.0072 &  0.1280 & {-0.0096} &  {-0.0139}        \\
          IP(III)  &  33.4930$^a$  & 33.4791 & -0.2143 & -0.2827 & -0.1909 & -0.2143 & -0.0744 & {-0.2175} &  {-0.0173}        \\
           IP(IV)  &  45.1418$^a$  & 45.1325 & -0.4016 & -0.5920 & -0.3528 & -0.4683 & -0.1474 & {-0.5123} &  { 0.0242}              \\
               EA  &  1.3895$^b$   &  1.3928 &  0.0097 &  0.0210 &  0.0144 &  0.0193 &  0.0417 & { 0.0175} &  { 0.0029}        \\
         \hline
  \multicolumn{3}{l}{ AMAD }        &     0.1326 &     0.1849 &     0.1173 &     0.1455 &     0.0978 &     {0.1547} &     {0.0119} \\
         \multicolumn{3}{l}{ LMAD } &      0.0157  &0.0165  &0.0142  &0.0149 &0.0891  & {0.0146}  & {0.0060}     \\
 \multicolumn{3}{l}{ MARE }        &     0.0050 &     0.0079 &     0.0053 &     0.0067 &     0.0111 &     {0.0066} &     {0.0008} \\ 
         \hline
         \hline
          \multicolumn{10}{l}{\footnotesize $^a$ Reference \cite{NIST_ASD}} \\ 
         \multicolumn{10}{l}{\footnotesize $^b$ Reference \cite{EA}}
    \end{tabular}
    \label{tab:si_atomic}
\end{table*}

\subsection{Phosphorus}
\begin{figure*}[!h]
\centering
\begin{subfigure}{0.5\textwidth}
\includegraphics[width=\textwidth]{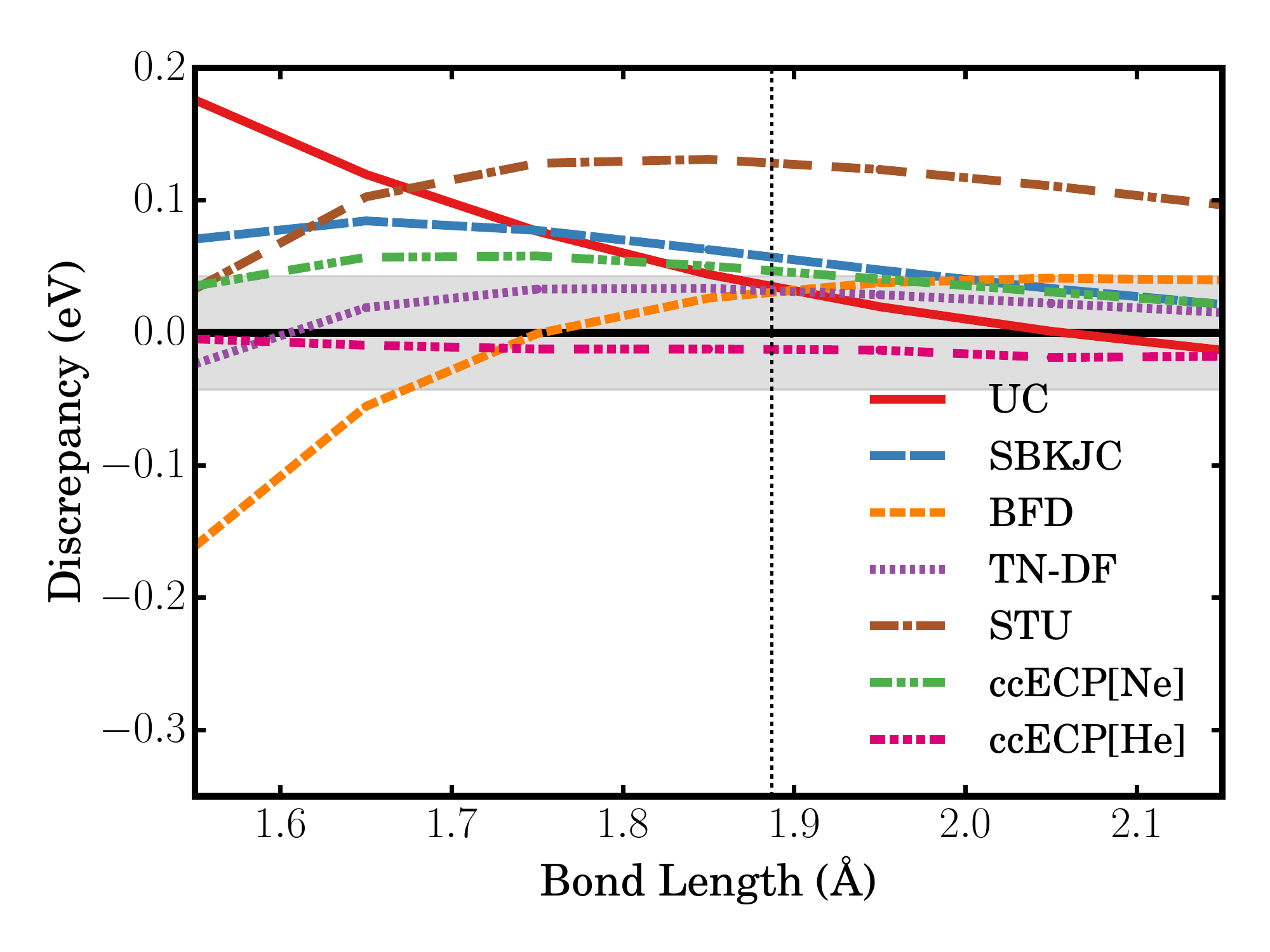}
\caption{P2 binding curve discrepancies}
\label{fig:P2}
\end{subfigure}%
\begin{subfigure}{0.5\textwidth}
\includegraphics[width=\textwidth]{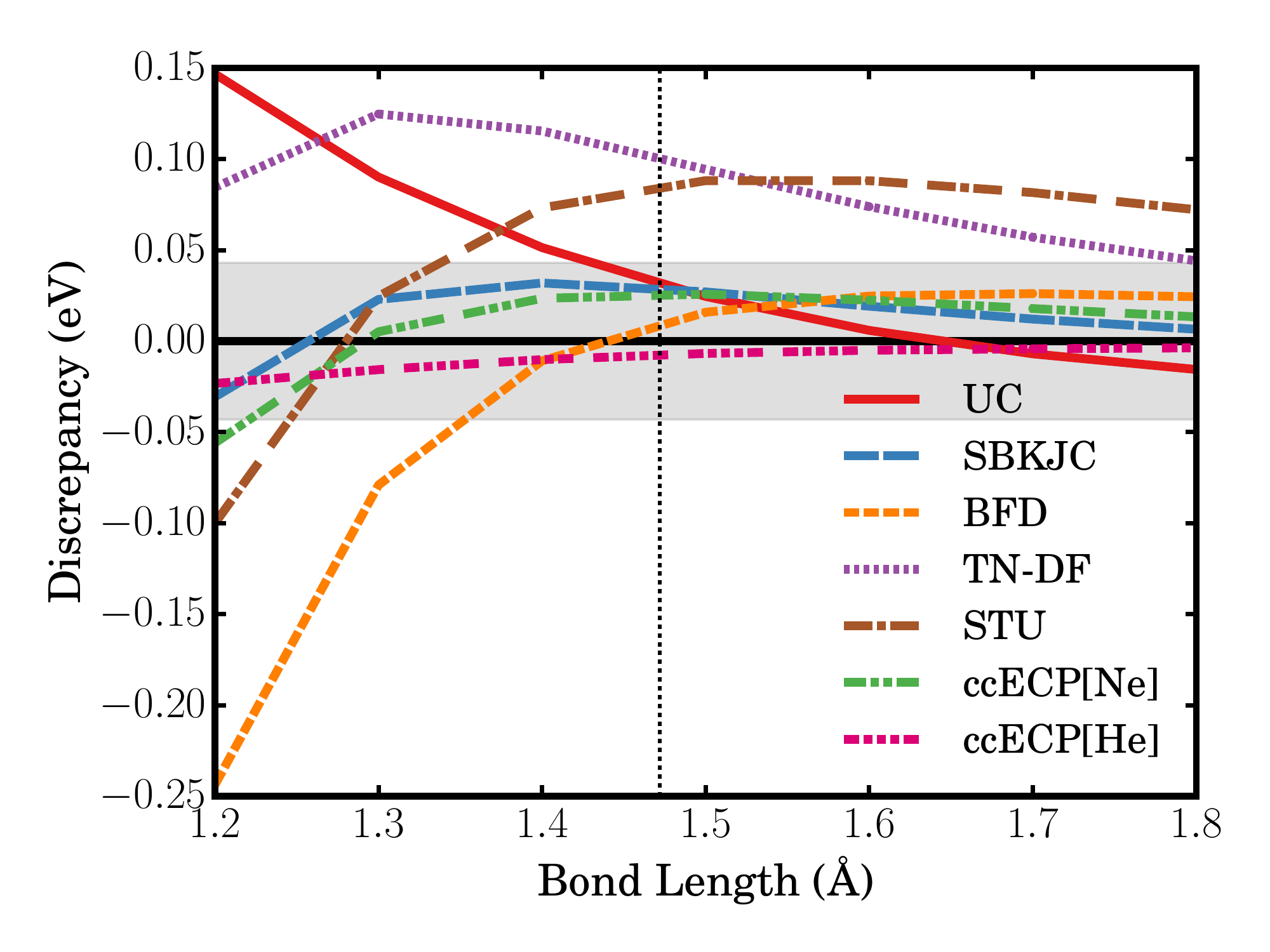}
\caption{PO binding curve discrepancies}
\label{fig:PO}
\end{subfigure}
\caption{Binding energy discrepancies for (a) P2 and (b) PO molecules in their ground states $^1\Sigma_g$ and $^2\Pi$, respectively. The binding curves are relative to the AE UCCSD(T) binding curve. The shaded region indicates a discrepancy of chemical accuracy in either direction. }
\label{fig:P_mols}
\end{figure*}

For phosphorous, the atomic and molecular results of our Ne-core and He-core ECPs are shared in Table \ref{tab:p_atomic} and Fig. \ref{fig:P_mols}, respectively.
For the case of our Ne-core partition, we observed that comparable accuracies of the atomic properties were only achievable with respect to the other core approximations.
However, significant improvement was again achieved for the case of our He-core partition.

For the case of P$_2$, both core partitions perform well with the Ne-core ECP obtaining maximum errors that are only marginally larger than chemical accuracy and the He-core ECP's errors being a factor of 2-3 smaller still.
The results from PO are also quite good for both partitions.
At equilibrium, both He- and Ne-core ECPs show errors no more than about {$0.03$} eV.
For our He-core ECP, this level of accuracy is maintained as the bond is compressed, however, the Ne-core ECP at these shorter separations begins to overbind. Note that our constructions compete in overall optimality with SBKJC while having 
smaller variational freedom in bounded potentials. In addition, the molecular results show rather nonsystematic behavior for STU ECPs that are different (underbinding) from the rest of the row (typically, overbinding).

\begin{table*}[h!]
    \centering
    \caption{All-electron (AE) UCCSD(T) ionization potentials and electron affinity of P along with the errors from uncorrelated core (UC) and ECPs. The uncontracted aug-cc-pCV5Z basis was used for all calculations. All values in eV. See Tab. V for further description.}
    \begin{tabular}{lrrrrrrrrrrr}
           \hline
           \hline
           \multicolumn{1}{l}{\multirow{2}{*}{Qty.}} & \multicolumn{1}{r}{\multirow{2}{*}{Exp.}}  & \multicolumn{1}{r}{\multirow{2}{*}{AE}} &  \multicolumn{7}{c}{Discrepancies from AE}  \\
         &  &  & \multicolumn{1}{r}{UC} & \multicolumn{1}{r}{SBKJC} &     \multicolumn{1}{r}{BFD} &   \multicolumn{1}{r}{TN-DF} &     \multicolumn{1}{r}{STU} & \multicolumn{1}{r}{ccECP[Ne]} & \multicolumn{1}{r}{ccECP[He]} \\
           \hline
                IP(I)  &  10.4867$^a$  &  10.5104 & -0.0114 &  0.0171 &  0.0138 &  0.0221 & -0.0054 & { 0.0252} &  { 0.0011}       \\
               IP(II)  &  19.7695$^a$  &  19.7562 & -0.0355 & -0.0116 & -0.0003 &  0.0019 &  0.0050 & {-0.0018} &  {-0.0109}      \\
              IP(III)  &  30.2026$^a$  &  30.1328 & -0.0677 & -0.0798 & -0.0535 & -0.0522 &  0.0121 & {-0.0729} &  {-0.0347}       \\
               IP(IV)  &  51.4439$^a$  &  51.4304 & -0.2527 & -0.4559 & -0.2187 & -0.3254 & -0.2051 & {-0.3577} &  {-0.0241}       \\
                IP(V)  &  65.0251$^a$  &  65.0181 & -0.4402 & -0.9174 & -0.3825 & -0.6667 & -0.4504 & {-0.7180} &  { 0.0348}      \\
                   EA  &   0.7465$^b$  &   0.7003 &  0.0055 & -0.0103 & -0.0260 & -0.0102 & -0.0350 & {-0.0114} &  { 0.0020}       \\
         \hline
 \multicolumn{3}{l}{ AMAD }        &     0.1355 &     0.2487 &     0.1158 &     0.1797 &     0.1188 &    {0.1978} &     {0.0179} \\ 
         \multicolumn{3}{l}{ LMAD }               &   0.0175  &0.0130  &0.0134  &0.0114  &0.0151    &     {0.0128} &    {0.0047} \\
 \multicolumn{3}{l}{ MARE }      &     0.0041 &     0.0071 &     0.0084 &     0.0058 &     0.0103   &     {0.0065} &    {0.0009} \\
         \hline
         \hline
          \multicolumn{10}{l}{\footnotesize $^a$ Reference \cite{NIST_ASD}} \\ 
         \multicolumn{10}{l}{\footnotesize $^b$ Reference \cite{EA}}
    \end{tabular}
    \label{tab:p_atomic}
\end{table*}

\subsection{Sulfur}
\begin{figure*}[!h]
\centering
\begin{subfigure}{0.5\textwidth}
\includegraphics[width=\textwidth]{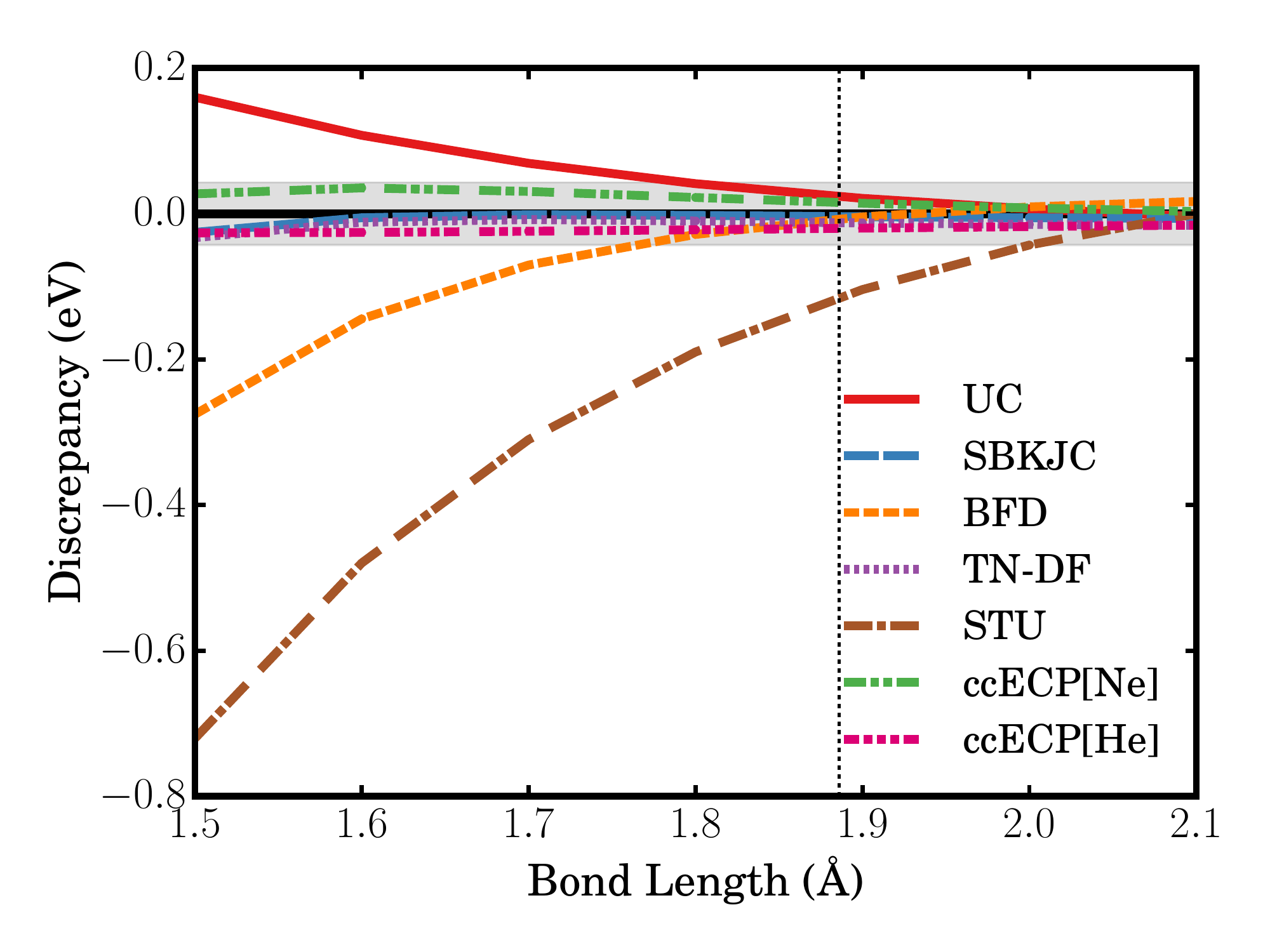}
\caption{S$_2$ binding curve discrepancies}
\label{fig:S2}
\end{subfigure}%
\begin{subfigure}{0.5\textwidth}
\includegraphics[width=\textwidth]{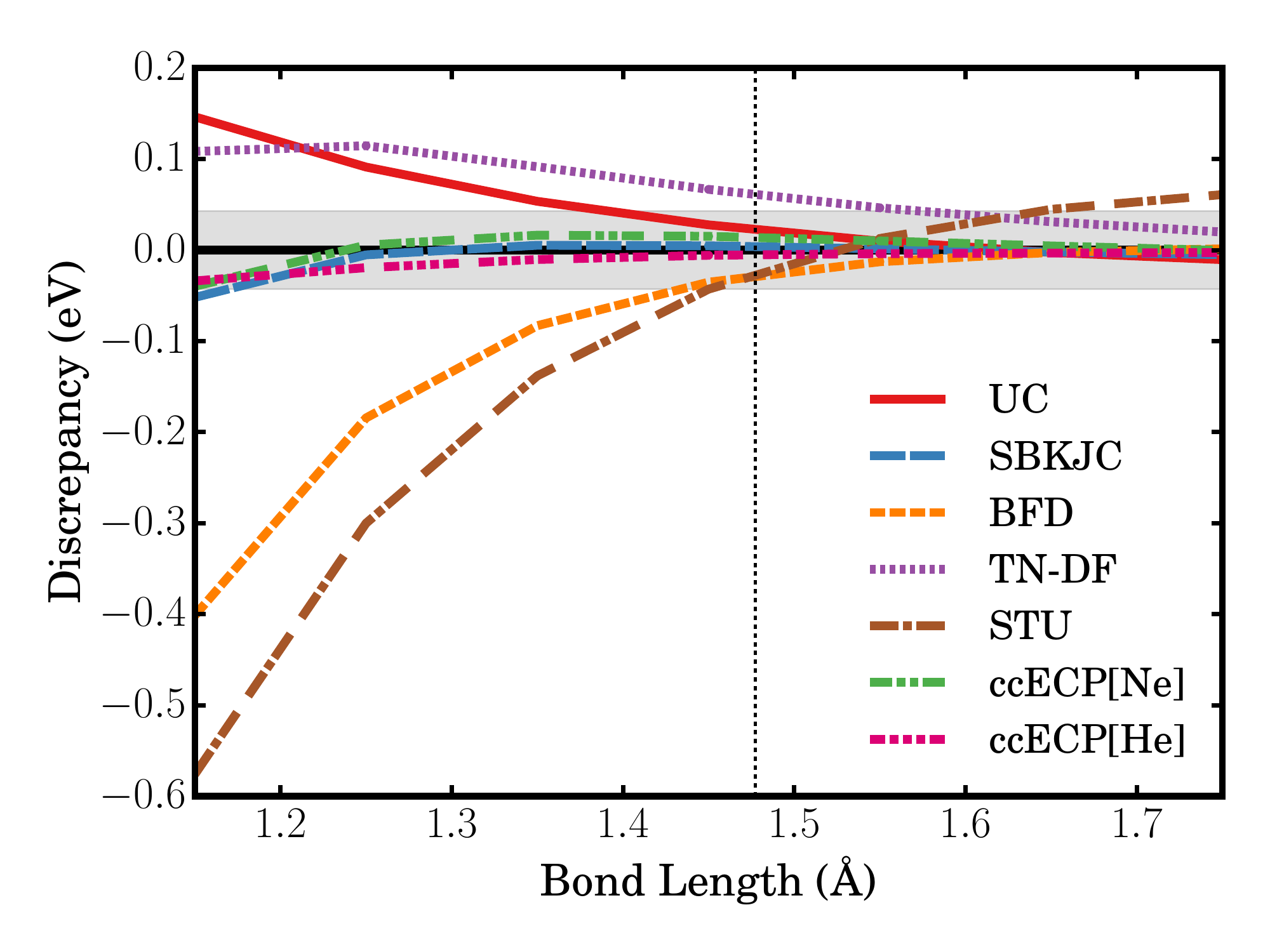}
\caption{SO binding curve discrepancies}
\label{fig:SO}
\end{subfigure}
\caption{Binding energy discrepancies for (a) S$_2$ and (b) SO molecules in their ground states $^3\Sigma_g$ and $^3\Sigma$, respectively. The binding curves are relative to the AE UCCSD(T) binding curve. The shaded region indicates a discrepancy of chemical accuracy in either direction. }
\label{fig:S_mols}
\end{figure*}

For sulfur, the atomic and molecular results of our Ne-core and He-core ECPs are shared in Table \ref{tab:s_atomic} and Fig. \ref{fig:S_mols}, respectively.
In this case, the accuracies of the Ne-core ECP at the UCCSD(T) level are rather good with errors within chemical accuracies throughout all bond lengths plotted for both S$_2$ and SO molecules.
And for our He-core ECP, we again see the benefit of this particular choice of core-valence partitioning where the atomic accuracies show large improvements over the other Ne-core approximations and {$\approx 0.02$-$0.03$} eV accuracy is achieved for both the sulfur dimer and oxide across all geometries plotted.

\begin{table*}[h!]
    \centering
    \caption{All-electron (AE) UCCSD(T) valence ionization potentials and electron affinity of S along with the errors from uncorrelated core (UC) and ECPs. The uncontracted aug-cc-pCV5Z basis was used for all calculations. All values in eV. See Tab. V for further description.}
    \begin{tabular}{lrrrrrrrrrrr}
           \hline
           \hline
           \multicolumn{1}{l}{\multirow{2}{*}{Qty.}} &  \multicolumn{1}{r}{\multirow{2}{*}{Exp.}} & \multicolumn{1}{r}{\multirow{2}{*}{AE}} &  \multicolumn{7}{c}{Discrepancies from AE}  \\
         &  &  & \multicolumn{1}{r}{UC} & \multicolumn{1}{r}{SBKJC} &     \multicolumn{1}{r}{BFD} &   \multicolumn{1}{r}{TN-DF} &     \multicolumn{1}{r}{STU} & \multicolumn{1}{r}{ccECP[Ne]} & \multicolumn{1}{r}{ccECP[He]} \\
          \hline
            IP(I)  &  10.3600$^a$           &       10.2999 &       -0.0166 &       -0.0338 &       -0.0583 &       -0.0335 &        0.0851 &    {-0.0397}  &    { 0.0028}  \\
           IP(II)  &  23.3378$^a$           &       23.3950 &       -0.0375 &        0.0231 &        0.0234 &        0.0098 &        0.2906 &    { 0.0194}  &    { 0.0001}  \\
          IP(III)  &  34.86$^a$\,\,\,\,\,\, &       34.8258 &       -0.0670 &       -0.0301 &        0.0010 &       -0.0431 &        0.3369 &    {-0.0388}  &    {-0.0202}  \\
           IP(IV)  &  47.222$^a$\,\,\,      &       47.2693 &       -0.1042 &       -0.1361 &       -0.0673 &       -0.1393 &        0.3210 &    {-0.1532}  &    {-0.0556}   \\
            IP(V)  &  72.5945$^a$           &       72.5882 &       -0.2887 &       -0.6879 &       -0.2331 &       -0.4965 &       -0.0172 &    {-0.5524}  &    {-0.0360}   \\
           IP(VI)  &  88.0529$^a$           &       88.0550 &       -0.4769 &       -1.1502 &       -0.3611 &       -0.8586 &       -0.4356 &    {-0.9956}  &    { 0.0532}  \\
               EA  &  2.0771$^b$            &        2.0500 &        0.0019 &        0.0020 &       -0.0276 &        0.0004 &        0.0326 &    {-0.0026}  &    { 0.0056}    \\
         \hline
       \multicolumn{3}{l}{ AMAD }       &     0.1418 &     0.2947 &     0.1103 &     0.2259 &     0.2170 &     { 0.2574} &     { 0.0248} \\ 
         \multicolumn{3}{l}{ LMAD }                &0.0187  &0.0196  &0.0364  &0.0146 &0.1361  & { 0.0206} & { 0.0028}           \\
 \multicolumn{3}{l}{ MARE }       &     0.0025 &     0.0045 &     0.0041 &     0.0035 &     0.0083 &    { 0.0042}  &  { 0.0008} \\
        
         \hline
         \hline
          \multicolumn{10}{l}{\footnotesize $^a$ Reference \cite{NIST_ASD}} \\ 
         \multicolumn{10}{l}{\footnotesize $^b$ Reference \cite{EA}}
    \end{tabular}
    \label{tab:s_atomic}
\end{table*}

\subsection{Chlorine}
\begin{figure*}[!h]
\centering
\begin{subfigure}{0.5\textwidth}
\includegraphics[width=\textwidth]{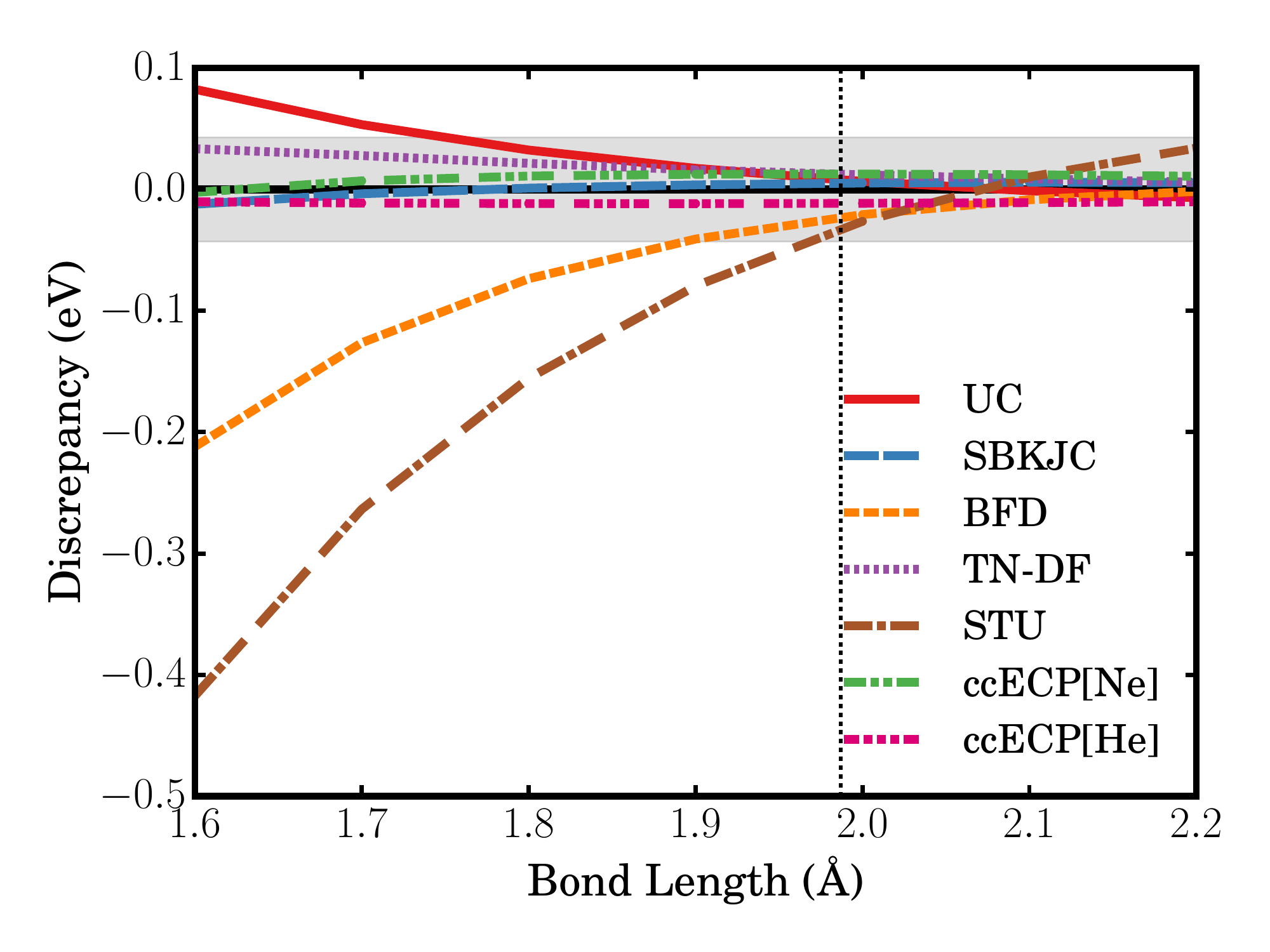}
\caption{Cl2 binding curve discrepancies}
\label{fig:Cl2}
\end{subfigure}%
\begin{subfigure}{0.5\textwidth}
\includegraphics[width=\textwidth]{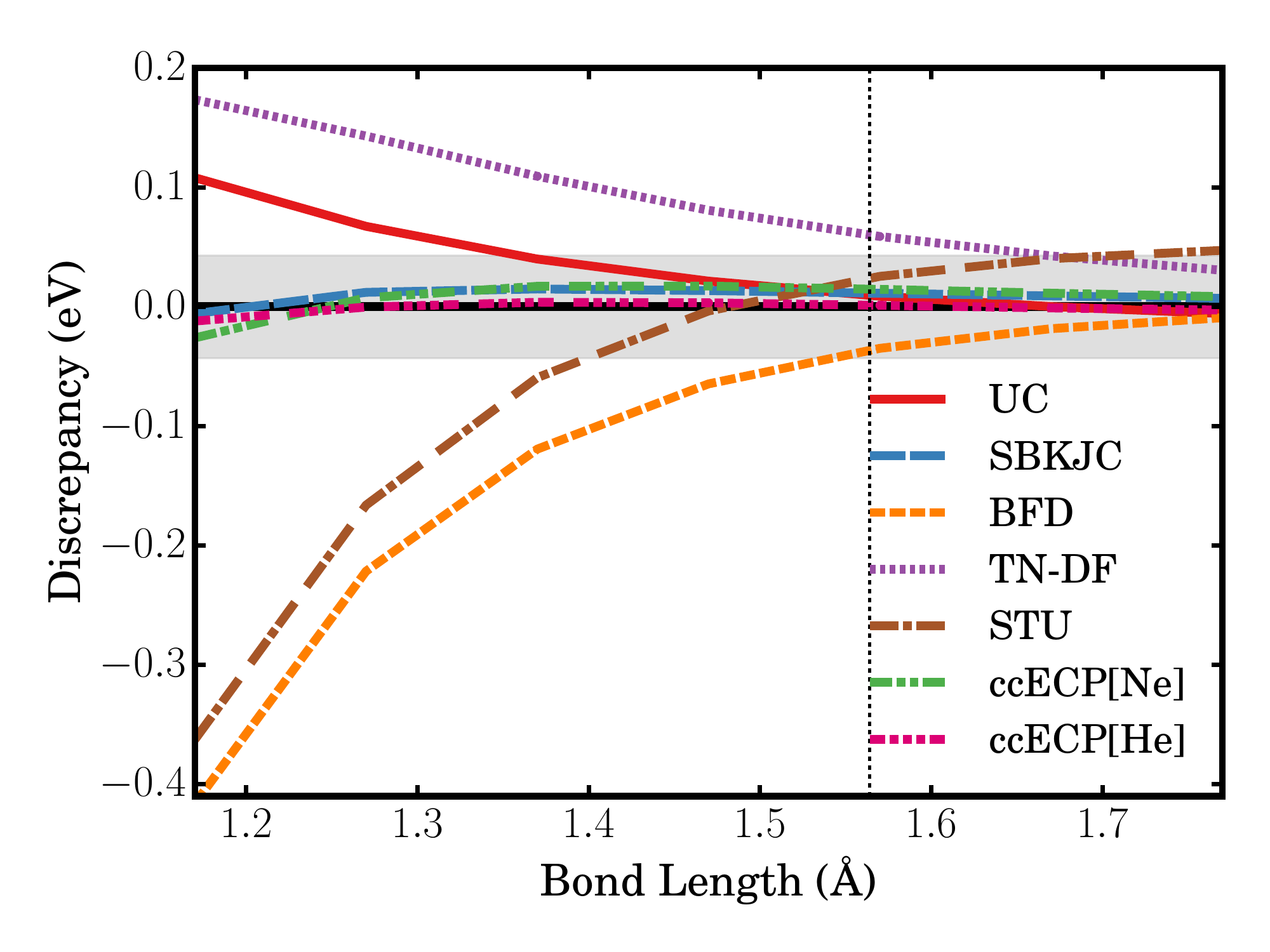}
\caption{ClO binding curve discrepancies}
\label{fig:ClO}
\end{subfigure}
\caption{Binding energy discrepancies for (a) Cl2 and (b) ClO molecules in their ground states $^1\Sigma_g$ and $^2\Pi$, respectively. The binding curves are relative to the AE UCCSD(T) binding curve. The shaded region indicates a discrepancy of chemical accuracy in either direction. }
\label{fig:Cl_mols}
\end{figure*}

For chlorine, the atomic and molecular results of our Ne-core and He-core ECPs are shared in table \ref{tab:cl_atomic} and figure \ref{fig:Cl_mols}, respectively.
For this case, we only observed a significant improvement of the ionization potentials {and} electron affinity for our He-core ECP -- its errors reaching no more than about $0.1$ eV even for very deeply ionized cases.

Our construction, however, led to He- and Ne-core ECPs that reproduce the all-electron Cl$_2$ and ClO potential energy surfaces quite well -- {where both large and small core ECPs remain well within chemical accuracy for the full range of geometries. 
To provide another option with lower spectral errors, given the larger errors of our Ne-core ECPs in deeply ionized cases, we also generated a Ne-core ECP that weighted the energy component of the objective function more heavily to reduce the spectral errors further at the cost of slightly larger errors in the accuracy of Cl$_2$ and ClO. We share this alternative ECP in the supplemental material.}

\begin{table*}[h!]
    \centering
    \caption{All-electron (AE) UCCSD(T) ionization potentials and electron affinity of Cl along with the errors from uncorrelated core (UC) and ECPS. The uncontracted aug-cc-pCV5Z basis was used for all calculations. All values in eV. See Tab. V for further description.}
    \begin{tabular}{lrrrrrrrrrrr}
           \hline
           \hline
           \multicolumn{1}{l}{\multirow{2}{*}{Qty.}} & \multicolumn{1}{r}{\multirow{2}{*}{Exp.}} & \multicolumn{1}{r}{\multirow{2}{*}{AE}} &  \multicolumn{7}{c}{Discrepancies from AE}  \\
         &  &  & \multicolumn{1}{r}{UC} & \multicolumn{1}{r}{SBKJC} &     \multicolumn{1}{r}{BFD} &   \multicolumn{1}{r}{TN-DF} &     \multicolumn{1}{r}{STU} & \multicolumn{1}{r}{ccECP[Ne]} & \multicolumn{1}{r}{ccECP[He]} \\
           \hline
      IP(I)                &  12.9676$^a$  &  12.9388 &           -0.0175 & -0.0082 & -0.0396 & -0.0210 &  0.1591 &   {-0.0121} &       { 0.0015}    \\
     IP(II)                &  23.8136$^a$  &  23.7309 &           -0.0401 & -0.0649 & -0.0860 & -0.0743 &  0.2194 &   {-0.0637} &         {-0.0103}    \\
    IP(III)                &  39.80$^a$\,\,\,\,\,\,  &  39.6980 & -0.0614 &  0.0090 &  0.0091 & -0.0126 &  0.4779 &   { 0.0234} &         {-0.0238}    \\
     IP(IV)                &  53.24$^a$\,\,\,\,\,\,  &  53.1811 & -0.0939 & -0.0807 & -0.0816 & -0.0952 &  0.5020 &   {-0.0655} &         {-0.0572}    \\
      IP(V)                &  67.68$^a$\,\,\,\,\,\,  &  67.6102 & -0.1328 & -0.2335 & -0.2471 & -0.2280 &  0.4487 &   {-0.2217} &         {-0.1069}           \\
     IP(VI)                &  96.94$^a$\,\,\,\,\,\,  &  96.8961 & -0.3139 & -0.9935 & -0.0873 & -0.6785 &  0.0301 &   {-0.7001} &         {-0.0441}           \\
    IP(VII)                &  114.2013$^a$  & 114.2079          & -0.5023 & -1.5507 & -0.2635 & -1.0900 & -0.4557 &   {-1.1858} &         { 0.0646}          \\
         EA                &  3.6127$^b$  &   3.6210            & -0.0007 &  0.0193 & -0.0153 &  0.0079 &  0.0781 &   {-0.0123} &         { 0.0064}           \\
                  \hline
     \multicolumn{3}{l}{ AMAD }        &     0.1453 &     0.3700 &     0.1037 &     0.2759 &     0.2964 &     {0.2856} &     {0.0394} \\ 
         \multicolumn{3}{l}{ LMAD }                &0.0194  &0.0308  &0.0470  &0.0344 &0.1522  &{0.0294}  & {0.0061}           \\
 \multicolumn{3}{l}{ MARE }       &     0.0020 &     0.0047 &     0.0024 &     0.0036 &     0.0094 &     {0.0037} &   {0.0008} \\ 
         \hline
         \hline
          \multicolumn{10}{l}{\footnotesize $^a$ Reference \cite{NIST_ASD}} \\ 
         \multicolumn{10}{l}{\footnotesize $^b$ Reference \cite{EA}}
    \end{tabular}
    \label{tab:cl_atomic}
\end{table*}

\subsection{Argon}
\begin{figure}[!h]
\centering
\caption{Binding energy discrepancies for the ArH$+$ molecule in its ground state $^1\Sigma$. The binding curves are relative to the AE UCCSD(T) binding curve. The shaded region indicates a discrepancy of chemical accuracy in either direction. }
\includegraphics[width=0.5\textwidth]{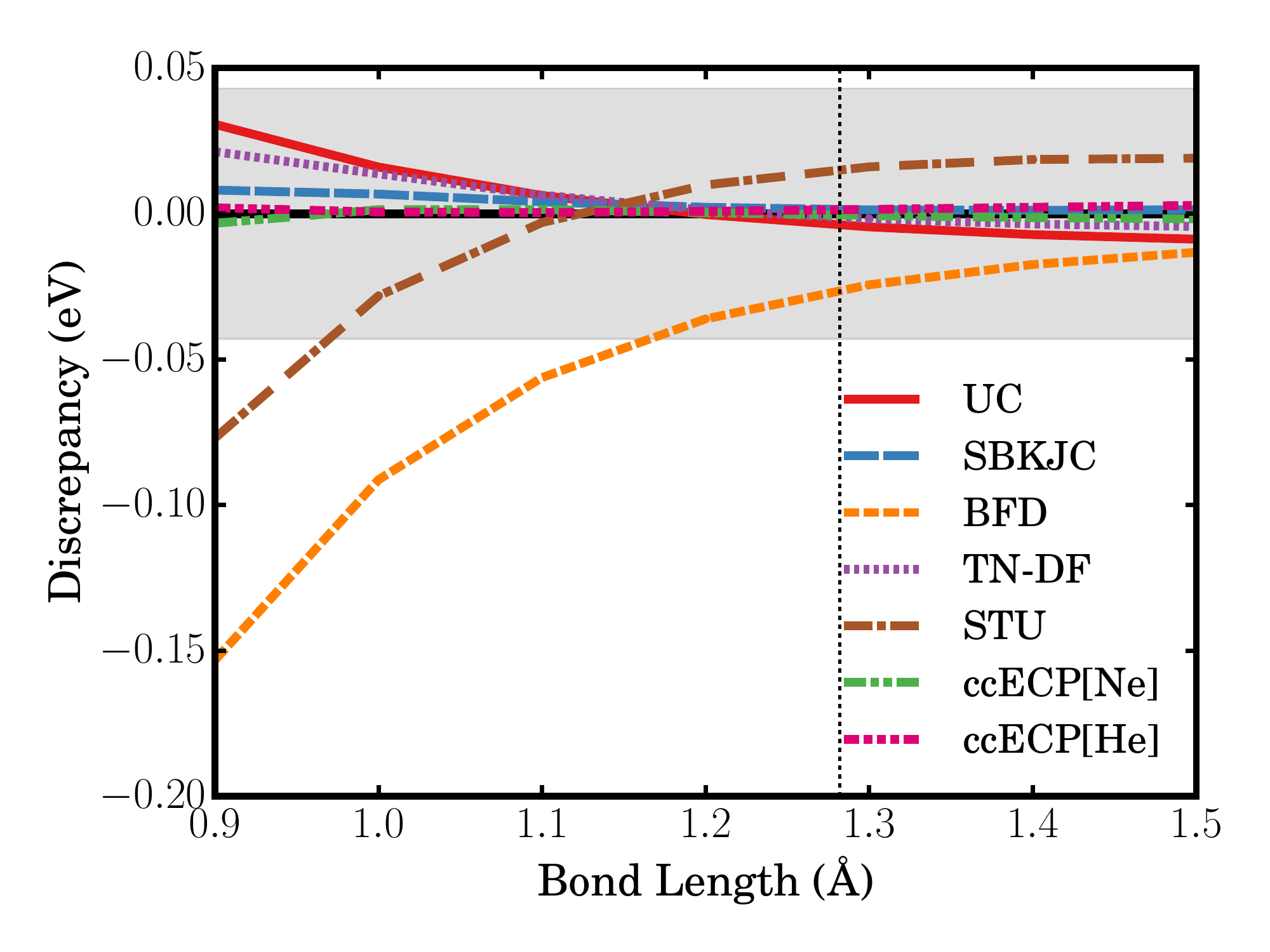}
\label{fig:ArH}
\end{figure}

For argon, the atomic and molecular results of our Ne-core and He-core ECPs are shared in Table \ref{tab:ar_atomic} and Fig. \ref{fig:ArH}, respectively.
With respect to the atomic properties, the Ne-core ECP performs well in this case with a total mean absolute deviation that is comparable to the all-electron uncorrelated core results.
The He-core again performs even better where the total mean absolute deviation is improved over our Ne-core ECP by another {$0.4$} eV.
To test the ECPs' performances in a molecular setting we calculated the errors from the all-electron UCCSD(T) ArH+ binding curve.
Both core partitions show high accuracies for this case where the Ne-core and He-core errors are no larger than {$\approx 0.01$ eV in magnitude}.
{Similar to chlorine, we also generated a Ne-core ECP that weighted the energy component of the objective function more heavily to reduce the spectral errors further at the cost of larger errors in the accuracy 
in the molecule. We share this alternative ECP in the supplemental material.}

\begin{table*}[h!]
    \centering
    \caption{All-electron (AE) UCCSD(T) ionization potentials for Ar along with the errors from uncorrelated core (UC) and ECPs. The uncontracted aug-cc-pCV5Z basis was used for all calculations. All values in eV. See Tab. V for further description.}
    \begin{tabular}{lrrrrrrrrrrr}
           \hline
           \hline
           \multicolumn{1}{l}{\multirow{2}{*}{Qty.}} & \multicolumn{1}{r}{\multirow{2}{*}{Exp.}} & \multicolumn{1}{r}{\multirow{2}{*}{AE}} &  \multicolumn{7}{c}{Discrepancies from AE}  \\
         &  &  & \multicolumn{1}{r}{UC} & \multicolumn{1}{r}{SBKJC} &     \multicolumn{1}{r}{BFD} &   \multicolumn{1}{r}{TN-DF} &     \multicolumn{1}{r}{STU} & \multicolumn{1}{r}{ccECP[Ne]} & \multicolumn{1}{r}{ccECP[He]} \\
           \hline
IP(I)                &  15.7596$^a$  &     15.7829             &   -0.0172    &   0.0146    &  -0.0156   &   -0.0016   &   -0.0116   &   {-0.00327}       &  { 0.0201}  \\
IP(II)               &  27.6297$^a$  &     27.6005             &   -0.0369    &  -0.0386    &  -0.0510   &   -0.0567   &   -0.0071   &   {-0.05349}       &  { 0.0135}  \\
IP(III)              &  40.735$^a$\,\,\,  &     40.6121        &   -0.0619    &  -0.1305    &  -0.1243   &   -0.1451   &   -0.0132   &   {-0.13884}       &  {-0.0024}  \\
IP(IV)               &  59.58$^a$\,\,\,\,\,\,  &     59.2804   &   -0.0835    &  -0.0581    &  -0.0243   &   -0.0929   &    0.2038   &   {-0.05261}       &  {-0.0183}  \\
IP(V)                &  74.84$^a$\,\,\,\,\,\,  &     74.7437   &   -0.1178    &  -0.2015    &  -0.1742   &   -0.2220   &    0.2207   &   {-0.18799}       &  {-0.0622}   \\
IP(VI)               &  91.290$^a$\,\,\,  &     91.1085        &   -0.1580    &  -0.4181    &  -0.4141   &   -0.4046   &    0.2016   &   {-0.39684}       &  {-0.1260}   \\
IP(VII)              &  124.41$^a$\,\,\,\,\,\,  &    124.3356  &   -0.3353    &  -1.3918    &   0.0652   &   -0.9451   &   -0.4787   &   {-1.13569}       &  {-0.1348}  \\
IP(VIII)             &  143.4567$^b$  &    143.4706            &   -0.5238    &  -2.0213    &  -0.0730   &   -1.3990   &   -0.9534   &   {-1.7643}       &  { 0.1849}  \\
         \hline
              \multicolumn{3}{l}{ AMAD }       &     0.1668 &     0.5343 &     0.1177 &     0.4084 &     0.2613 &     {0.4666} &     {0.0703} \\ 
                   \multicolumn{3}{l}{ LMAD }        &0.0271  &0.0266  &0.0333  &0.0291  &0.0094  &{0.0284} &{0.0168}                  \\
 \multicolumn{3}{l}{ MARE }       &     0.0019 &     0.0049 &     0.0018 &     0.0040 &     0.0026 &     {0.0043} &    {0.0008} \\
         \hline
         \hline
          \multicolumn{10}{l}{\footnotesize $^a$ Reference \cite{NIST_ASD}} \\ 
         \multicolumn{10}{l}{\footnotesize $^b$ Reference \cite{EA}}
    \end{tabular}
    \label{tab:ar_atomic}
\end{table*}

\subsection{Molecular binding parameters, total energies and core radii}

We compiled all the results for molecular
equilibrium parameters into the Tab.
\ref{tab:global_mads}. The  
statistics provide useful information for an overall performance of the presented constructions as well as comparison with 
previously published tables. We have achieved rather significant improvements in the molecular properties due to the more flat discrepancy
curves than other ECPs.
In general, the ECPs have a tendency to overbind for short bonds and the cause is the missing core charge and the related core-valence repulsion. Interestingly
enough, except for Al and Si, we were able to 
keep these biases typically below 0.05 eV 
with mildly increased deviations (0.1 eV)
at very small bond length for Mg. 
Clearly, minor improvements might still be possible although we expect that one would encounter diminishing returns unless the ECP form is further generalized. On the other hand, He-core framework clearly eliminates these deficiencies and provide very high accuracy across the whole row. Finally, we list total HF and correlation energies from UCCSD(T) as extrapolated to the complete basis set limit in Table \ref{tab:totals} along with the core radii of our ccECPs. 
While HF energies are saturated to about 0.1 mHa or better in all cases, 
for Ne-core we estimate extrapolation errors on correlation energies to be within about 1 mHa.
For He-core correlation energies we guess underestimation by about 3 mHa for Na and up to about 6 mHa for Ar, showing limitations of 5Z basis set and excitations up to perturbational triples to fully capture the many-body effects of 16 electrons. 
This estimate comes from an approximate analysis of core-core, 
core-valence and valence-valence correlation energy components in all-electron calculations \cite{ranasinghe2015} with an added
correction from Dolg \cite{Dolg:1996cpl} 
that reflects increase in correlations in the ECP setting.
The core radii{, $r_l$,} of each ccECP is taken as the radial distance from the origin at which {the channel's full potential agrees} to within $10^{-5}$ Ha of the bare Coulomb potential, $-Z_{\rm eff}/r$.
{Additionally, we share the cutoffs, $r_{l,{\rm nl}}$, of each channel's non-local potential which we define as the point where the non-local potential drops below $10^{-5}$ Ha.
The non-local raddi have implications primarily related to efficiency of QMC calculations as further explained elsewhere \cite{CeperleyMitas}.}
Additional data can be found also in Supplementary Information.

\begin{table*}[ht!]
    \centering
	\caption{Mean absolute deviations of discrepancies of binding parameters of all molecules considered in this work at equilibrium ($D_e$, $r_e$ and $\omega_e$) and near the all-electron dissociation threshold, $D_{\mathrm{diss}} (\lessapprox 0.05$ \AA), at short bond lengths for our ECPs and previous constructions with respect to all-electron UCCSD(T) calculations. Errors in parenthesis  such as ``2.4(4)'' denote ``2.4$\pm$ {0.4}'' and 
    correspond to deviations of Morse
    potential fits to binding curves.}
    \begin{tabular}{c | ccccccc}
	\hline
	\hline
	                          &          UC &        SBKJC &   BFD &   TN-DF &  STU & ccECP[Ne] &  ccECP[He] \\
	\hline
	 $D_e$      (eV/$10^2$)     &    {1.9(4)} &   {2.4(3)} &  {3.2(4)}  &   {5.1(3)} &    {5.8(4)}  &  {2.0(4)} &    {0.8(4)} \\
     $\omega_e$ (cm$^{-1}$)     &    {6(3)} &     {14(3)}  &   {9(3)}  &   {23(3)} &     {12(3)} &     {11(3)} &     {1(3)} \\
	 $r_e$      (m\AA)    &    {17(2)} &     {10(2)} &    {15(2)}  &    {12(2)} &     {18(2)} &     {3(2)} &     {1(2)}  \\
	 $D_{\mathrm{diss}}$      (eV/$10^2$)  &  {16.9}  &   {16.9}  &   {31.9}   &  {16.5}   &  {38.9}  &   {9.2}  &  {1.4} \\
	 \hline
	 \hline
    \end{tabular}
    \label{tab:global_mads}
\end{table*}

\begin{table*}
{
\centering
    \caption{Total energies of our ccECPs with Ne (LC) 
    and He (SC) cores in their neutral ground states along with their core radii for each angular momentum channel -- the values $r_l$ are taken as the distance which the channel's full potential agrees with the bare Coloumb potential to within $10^{-5}$ Ha while $r_{l,\mathrm{nl}}$ is the distance at which the channel's non-local potential drops below $10^{-5}$ Ha. Both RHF/ROHF and UCCSD(T) correlation energies are extrapolated to the CBS limit using the uncontracted aug-cc-pwCV\{T,Q,5\}Z bases. Energies given in Ha. Radii given in \AA.}

\begin{tabular}{r|ccccc|cccccc}
\hline\hline
    Atom &  $r_{\rm s}$ & $r_{\rm p}$ & $r_{\rm d}$ & $r_{\rm s,nl}$ & $r_{\rm p,nl}$  & HF & Corr. \\
\hline
Na(LC) & 1.648 & 2.009  & 1.464 & 1.652  & 2.009  & -0.186203  &   \\    
Na(SC) & 0.675 & 0.675  &  & 0.543  &    & -47.357015 & -0.325551  \\    
\hline
Mg(LC) & 1.578 & 1.838  & 1.232 & 1.578  & 1.838   & -0.788370 & -0.035077  \\
Mg(SC) & 0.625 & 0.625  &  & 0.480  &    & -62.927451 & -0.363681  \\
\hline
Al(LC) & 1.406 & 1.633  & 1.135 & 1.406  & 1.633   & -1.876999 & -0.059869  \\
Al(SC) & 0.591 & 0.591  &  & 0.431  &    & -80.993054 & -0.395813  \\
\hline
Si(LC) & 1.273 & 1.427  & 1.006 & 1.273  & 1.427   & -3.672423 & -0.088666  \\
Si(SC) & 0.564 & 0.564  &  & 0.387  &    & -101.625857 & -0.428232  \\
\hline
 P(LC) & 1.173 & 1.278  & 0.925 & 1.173  & 1.278   & -6.340917 & -0.117781  \\
 P(SC) & 0.508 & 0.508  &  & 0.354  &    & -125.258405 & -0.461752  \\
\hline
 S(LC) & 1.085 & 1.165  & 0.867 & 1.085  & 1.165   & -9.918156 & -0.177994  \\
 S(SC) & 0.471 & 0.471  &  & 0.329  &    & -151.916385 & -0.525628  \\
\hline
Cl(LC) & 1.015 & 1.068  & 0.807 & 1.015  & 1.068   & -14.689386 & -0.235794  \\
Cl(SC) & 0.422 & 0.422  &  & 0.303  &    & -181.612402 & -0.587434  \\
\hline
Ar(LC) & 0.950 & 1.004  & 0.795 & 0.950  & 1.004   & -20.779601 & -0.289756  \\
Ar(SC) & 0.418 & 0.418  &  & 0.283  &    & -214.891481 & -0.642806  \\
\hline
\hline
\end{tabular}
\label{tab:totals}
}
\end{table*}

\section{Conclusions}

In this work, we offer several new advances and insights into the construction of correlation consistent ECPs and apply this to the construction of potentials for the 2nd row elements. 

In certain aspects, the 2nd row elements appear to be somewhat more complicated than the 1st row even though the corresponding valence spaces are very similar and the total and kinetic energies are lower (by about 20\% when compared with the 1st row) indicating thus smoother electronic densities. 
Intuitively, therefore, accurately solving systems of 2nd row elements might be considered ``easier" than, say, describing the deeper lying and more localized p-subshells in systems containing 1st row elements. 
However, the small number of valence electrons combined with larger core sizes generate larger biases than in the 1st row. In particular, the core-valence effects in correlation (important for spectra) compete with the absence of electrostatic core-valence interactions (important for short bonds) and therefore these tendencies are more difficult to reconcile.   
We found that the fits with many-body spectra only had limited accuracy but when we included more of the mean-field spatial information we were able to to reach the most acceptable compromises. 

We were lead to the particular construction scheme introduced in this work after analyzing the behavior of the one-particle HF energies together with the correlation energies. 
Not too surprisingly, we found that the correlation energies for particular states and for a broad variety of existing ECPs with different parameterizations were essentially constant -- the changes were on the order of $0.01$-$0.02$ eV. This appears consistent with the key tenet of effective 
core potential theory that implies that the correlation will faithfully ``follow" the constructed effective valence operators. 
However, we found this to be correct
essentially only up to about 0.1 eV accuracy
for atomic spectra. As soon as higher accuracy was sought and molecules were considered, the interplay between various energy components became more complicated and required refinements.

The testing of transferability proved to be very revealing.
In particular, for the aluminum oxide and silicon oxide molecules at short bond lengths, we see a significant departure from the assumed accuracy standards of $\approx$ 0.05 eV for both ours and existing ECPs.
We traced the deficiency to the missing electron core tails and likely core-valence effects which are absent in the ECP and would need to be included to reach higher accuracies. 
Similar problems have been noticed before, see for example, Ref.\cite{oganov2003}. Here, we present a systematic view on this issue for Ne-core ECPs within polar bonds at short separations and, in addition, we point out that elimination of this issue might require more elaborated solutions.
We have tested the inclusion of classical effective repulsive  charges that would correspond to the electronic core-core interactions that could reasonably restore the desired accuracy. 
In order to introduce this in a more thought-through fashion, we leave this aspect for future studies.
This is a notable point that 
can affect many oxide calculations\cite{oganov2003} such as Al and Si perovskites
 at high pressures  (for example, these are important 
components of the Earth's mantle). 
Due to the importance of oxides, it is highly desirable to probe for these particular aspects in calculations of solids and other systems.

The atomic spectra and Tab. \ref{tab:global_mads} show that apart from the inherent bias for a few short polar bonds our constructions show consistent accuracy in both atomic and molecular 
calculations and offer close to an optimal balance overall.
Our results, somewhat unexpectedly, are on par with (or in a few cases significantly improve upon) the ECPs from the SBKJC table. Except for the Na atom (which is a bit special having only one valence electron), SBKJC ECPs appear to represent perhaps the most systematic consistency among the tested previous constructions. Note that the SBKJC ECPs 
are not bound at the nucleus,
diverging as $1/r^2$ in the repulsive channels and as 
$-Z_{\rm eff}/r$ in the local channel. 
Due to our choice of the ccECP form with bounded potentials and minimized radial extent of the nonlocal terms (although not explicitly enforced) for several elements we hit a comparably optimal point as has been achieved in SBKJC table before. In some cases, we were able to actually find significantly better compromises. As mentioned before, we note 
that Ne-core ECPs for
atoms such as Na and Mg with one or two
valence electrons might prove to have only limited 
applicability in some bonding environments where shallow core states contribute to the chemical effects (for example, 
in compounds at high pressures).  

It is also clear that core polarization and relaxation potentials \cite{shirley1991,martin:prb1993,muller1984a,muller1984b} can alleviate some of the mentioned deficiencies. In this study, however, we wanted to see the limits of the simplest
ECP form before employing new terms. In addition, oxide molecules for elements in the middle might be difficult to ``fix'' since the root of the problem appears to be the missing repulsive (essentially classical) charge from core electrons and this might require a more general and more extensive modification of the ECP form.

On the other hand, 
since we report here on the existing limits of the Ne-core ECPs in their current forms, we have decided to provide another option for high accuracy calculations that offer a significant boost in accuracy and eliminate many compromises mentioned above.  
To this end, we constructed high accuracy He-core ECPs that significantly decrease the discrepancies with all-electron relativistic calculations. The gain is by an order of magnitude or more, with typical errors of 0.001-0.005  eV on the atomic side and $\approx$ 0.01 eV for molecular systems. 
Obviously, these operators imply additional cost that
is most pronounced in stochastic methods 
due to the increase of total energies and fluctuations. Note that some of that is compensated from
much smaller effective core radius of nonlocal terms \cite{CeperleyMitas}. 

Overall, we believe that the constructed ECPs provide both new insights as well as practical new options for high accuracy valence-only calculations. 
Supplementary material contains further information including very accurate CCSD(T) total energies extrapolated to the complete basis set limit (better than 1 mHa accuracy) for atomic states and relativistic CCSD(T) molecular potential energy surfaces and binding parameters from all-electron calculations as well as the discrepancies from all core approximations considered in this work -- experimental binding parameters are also included for comparison. All ccECPs and basis sets are accessible at the website \cite{website}.  Besides this data, the presented analysis offers a clear comparison between existing ECPs
and should open new perspectives for further advances in this important research area.

Input and output files for the calculations performed in this study are available and maintained at the Materials Data Facility \cite{data_doi}.

\section*{Supplementary Material}
See supplementary material for additional calculated properties of our ccECPs including total energy components of various atomic states and the potential energy surfaces of the molecules considered in this work along with experimental binding properties.

\section*{Acknowledgements}

We would like to thank {P. R. C. Kent} for reading the manuscript and for {helpful} suggestions.
{We are also grateful for the thorough reading of the manuscript and insightful suggestions from one of the referees which lead to significant revisions and improvements to the manuscript.}
The majority of this work (development of the methods, calculations, tests, and writing of the paper) has been supported by the U.S. Department of Energy, Office of Science, Basic Energy Sciences, Materials Sciences and Engineering Division, as part of the Computational Materials Sciences Program and Center for Predictive Simulation of Functional Materials.
The initial theoretical and conceptual considerations were supported by ORNL/UT Batelle, LLC, subcontract N. 4000144475.

The calculations for this work were performed mostly at Sandia National Laboratories, while some of the calculation have been carried out at TACC under XSEDE.

Sandia National Laboratories is a multimission laboratory managed and operated by National Technology and Engineering Solutions of Sandia LLC, a wholly owned subsidiary of Honeywell International Inc. for the U.S. Department of Energy's National Nuclear Security Administration under contract DE-NA0003525.

\bibliography{main}

\begin{thebibliography}{45}
\expandafter\ifx\csname natexlab\endcsname\relax\def\natexlab#1{#1}\fi
\expandafter\ifx\csname bibnamefont\endcsname\relax
  \def\bibnamefont#1{#1}\fi
\expandafter\ifx\csname bibfnamefont\endcsname\relax
  \def\bibfnamefont#1{#1}\fi
\expandafter\ifx\csname citenamefont\endcsname\relax
  \def\citenamefont#1{#1}\fi
\expandafter\ifx\csname url\endcsname\relax
  \def\url#1{\texttt{#1}}\fi
\expandafter\ifx\csname urlprefix\endcsname\relax\def\urlprefix{URL }\fi
\providecommand{\bibinfo}[2]{#2}
\providecommand{\eprint}[2][]{\url{#2}}

\bibitem[{\citenamefont{{Koloren\v c} and Mitas}(2011)}]{kolorenc2011}
\bibinfo{author}{\bibfnamefont{J.}~\bibnamefont{{Koloren\v c}}}
  \bibnamefont{and} \bibinfo{author}{\bibfnamefont{L.}~\bibnamefont{Mitas}},
  \bibinfo{journal}{Rep. Prog. Phys.} \textbf{\bibinfo{volume}{74}},
  \bibinfo{pages}{026502} (\bibinfo{year}{2011}).

\bibitem[{\citenamefont{Wagner and Ceperley}(2016)}]{wagner2016}
\bibinfo{author}{\bibfnamefont{L.~K.} \bibnamefont{Wagner}} \bibnamefont{and}
  \bibinfo{author}{\bibfnamefont{D.~M.} \bibnamefont{Ceperley}},
  \bibinfo{journal}{Rep. Prog. Phys.} \textbf{\bibinfo{volume}{79}},
  \bibinfo{pages}{094501} (\bibinfo{year}{2016}).

\bibitem[{\citenamefont{Dubecky et~al.}(2016)\citenamefont{Dubecky, Mitas, and
  Jurecka}}]{dubecky2016}
\bibinfo{author}{\bibfnamefont{M.}~\bibnamefont{Dubecky}},
  \bibinfo{author}{\bibfnamefont{L.}~\bibnamefont{Mitas}}, \bibnamefont{and}
  \bibinfo{author}{\bibfnamefont{P.}~\bibnamefont{Jurecka}},
  \bibinfo{journal}{Chem. Rev.} \textbf{\bibinfo{volume}{116}},
  \bibinfo{pages}{5188} (\bibinfo{year}{2016}).

\bibitem[{\citenamefont{Needs et~al.}(2010)\citenamefont{Needs, Towler,
  Drummond, and {L\' opez R\'\i os}}}]{needs2010}
\bibinfo{author}{\bibfnamefont{R.~J.} \bibnamefont{Needs}},
  \bibinfo{author}{\bibfnamefont{M.~D.} \bibnamefont{Towler}},
  \bibinfo{author}{\bibfnamefont{N.~D.} \bibnamefont{Drummond}},
  \bibnamefont{and} \bibinfo{author}{\bibfnamefont{P.}~\bibnamefont{{L\' opez
  R\'\i os}}}, \bibinfo{journal}{J. Phys.: Condens. Matter}
  \textbf{\bibinfo{volume}{22}}, \bibinfo{pages}{023201}
  (\bibinfo{year}{2010}).

\bibitem[{\citenamefont{Booth et~al.}(2013)\citenamefont{Booth, Gruneis,
  Kresse, and Alavi}}]{booth2013}
\bibinfo{author}{\bibfnamefont{G.}~\bibnamefont{Booth}},
  \bibinfo{author}{\bibfnamefont{A.}~\bibnamefont{Gruneis}},
  \bibinfo{author}{\bibfnamefont{G.}~\bibnamefont{Kresse}}, \bibnamefont{and}
  \bibinfo{author}{\bibfnamefont{A.}~\bibnamefont{Alavi}},
  \bibinfo{journal}{Nature} \textbf{\bibinfo{volume}{493}},
  \bibinfo{pages}{365} (\bibinfo{year}{2013}).

\bibitem[{\citenamefont{Hamann et~al.}(1979)\citenamefont{Hamann, Schluter, and
  Chiang}}]{Hamann}
\bibinfo{author}{\bibfnamefont{D.~R.} \bibnamefont{Hamann}},
  \bibinfo{author}{\bibfnamefont{M.}~\bibnamefont{Schluter}}, \bibnamefont{and}
  \bibinfo{author}{\bibfnamefont{C.}~\bibnamefont{Chiang}},
  \bibinfo{journal}{Phys. Rev. Lett.} \textbf{\bibinfo{volume}{43}},
  \bibinfo{pages}{1494} (\bibinfo{year}{1979}).

\bibitem[{\citenamefont{Bachelet et~al.}(1984)\citenamefont{Bachelet, Hamann,
  and Schluter}}]{bachelet1984}
\bibinfo{author}{\bibfnamefont{G.~B.} \bibnamefont{Bachelet}},
  \bibinfo{author}{\bibfnamefont{D.~R.} \bibnamefont{Hamann}},
  \bibnamefont{and} \bibinfo{author}{\bibfnamefont{M.}~\bibnamefont{Schluter}},
  \bibinfo{journal}{Phys. Rev. B} \textbf{\bibinfo{volume}{29}},
  \bibinfo{pages}{2309} (\bibinfo{year}{1984}).

\bibitem[{\citenamefont{Troullier and Martins}(1991)}]{tm:prb1991}
\bibinfo{author}{\bibfnamefont{N.}~\bibnamefont{Troullier}} \bibnamefont{and}
  \bibinfo{author}{\bibfnamefont{J.~L.} \bibnamefont{Martins}},
  \bibinfo{journal}{Phys. Rev. B} \textbf{\bibinfo{volume}{43}},
  \bibinfo{pages}{1993} (\bibinfo{year}{1991}).

\bibitem[{\citenamefont{Vanderbilt}(1990)}]{vanderbilt1990}
\bibinfo{author}{\bibfnamefont{D.}~\bibnamefont{Vanderbilt}},
  \bibinfo{journal}{Phys. Rev. B} \textbf{\bibinfo{volume}{41}},
  \bibinfo{pages}{7892(R)} (\bibinfo{year}{1990}).

\bibitem[{\citenamefont{Hamann}(2013)}]{Hamann2013}
\bibinfo{author}{\bibfnamefont{D.~R.} \bibnamefont{Hamann}},
  \bibinfo{journal}{Phys. Rev. B} \textbf{\bibinfo{volume}{88}},
  \bibinfo{pages}{085117} (\bibinfo{year}{2013}).

\bibitem[{\citenamefont{Goedecker et~al.}(1996)\citenamefont{Goedecker, Teter,
  , and Hutter}}]{goedecker1996}
\bibinfo{author}{\bibfnamefont{S.}~\bibnamefont{Goedecker}},
  \bibinfo{author}{\bibfnamefont{M.}~\bibnamefont{Teter}}, , \bibnamefont{and}
  \bibinfo{author}{\bibfnamefont{J.}~\bibnamefont{Hutter}},
  \bibinfo{journal}{Phys. Rev. B} \textbf{\bibinfo{volume}{54}},
  \bibinfo{pages}{1703} (\bibinfo{year}{1996}).

\bibitem[{\citenamefont{Willand et~al.}(2013)\citenamefont{Willand, Kvashnin,
  Genovese, Vazquez-Mayagoitia, Deb, Sadeghi, Deutsch, and
  Goedecker}}]{goedecker2013}
\bibinfo{author}{\bibfnamefont{A.}~\bibnamefont{Willand}},
  \bibinfo{author}{\bibfnamefont{Y.~O.} \bibnamefont{Kvashnin}},
  \bibinfo{author}{\bibfnamefont{L.}~\bibnamefont{Genovese}},
  \bibinfo{author}{\bibfnamefont{A.}~\bibnamefont{Vazquez-Mayagoitia}},
  \bibinfo{author}{\bibfnamefont{A.~K.} \bibnamefont{Deb}},
  \bibinfo{author}{\bibfnamefont{A.}~\bibnamefont{Sadeghi}},
  \bibinfo{author}{\bibfnamefont{T.}~\bibnamefont{Deutsch}}, \bibnamefont{and}
  \bibinfo{author}{\bibfnamefont{S.}~\bibnamefont{Goedecker}},
  \bibinfo{journal}{J. Chem. Phys.} \textbf{\bibinfo{volume}{138}},
  \bibinfo{pages}{104109} (\bibinfo{year}{2013}).

\bibitem[{\citenamefont{Pickett}(1989)}]{Pickett}
\bibinfo{author}{\bibfnamefont{W.~E.} \bibnamefont{Pickett}},
  \bibinfo{journal}{Comp. Phys. Rep.} \textbf{\bibinfo{volume}{9}},
  \bibinfo{pages}{115} (\bibinfo{year}{1989}).

\bibitem[{\citenamefont{Dolg and Cao}(2012)}]{DolgCao}
\bibinfo{author}{\bibfnamefont{M.}~\bibnamefont{Dolg}} \bibnamefont{and}
  \bibinfo{author}{\bibfnamefont{X.}~\bibnamefont{Cao}},
  \bibinfo{journal}{Chem. Rev.} \textbf{\bibinfo{volume}{112}},
  \bibinfo{pages}{403} (\bibinfo{year}{2012}).

\bibitem[{\citenamefont{Shulenburger and
  Mattsson}(2013)}]{Shulenburger:2013prb}
\bibinfo{author}{\bibfnamefont{L.}~\bibnamefont{Shulenburger}}
  \bibnamefont{and} \bibinfo{author}{\bibfnamefont{T.~R.}
  \bibnamefont{Mattsson}}, \bibinfo{journal}{Phys. Rev. B}
  \textbf{\bibinfo{volume}{88}}, \bibinfo{pages}{245117}
  (\bibinfo{year}{2013}).

\bibitem[{\citenamefont{Foyevtsova et~al.}(2014)\citenamefont{Foyevtsova,
  Krogel, Kim, Kent, Dagotto, and Reboredo}}]{foyevtsova2014}
\bibinfo{author}{\bibfnamefont{K.}~\bibnamefont{Foyevtsova}},
  \bibinfo{author}{\bibfnamefont{J.~T.} \bibnamefont{Krogel}},
  \bibinfo{author}{\bibfnamefont{J.}~\bibnamefont{Kim}},
  \bibinfo{author}{\bibfnamefont{P.}~\bibnamefont{Kent}},
  \bibinfo{author}{\bibfnamefont{E.}~\bibnamefont{Dagotto}}, \bibnamefont{and}
  \bibinfo{author}{\bibfnamefont{F.~A.} \bibnamefont{Reboredo}},
  \bibinfo{journal}{Phys. Rev. X} \textbf{\bibinfo{volume}{4}},
  \bibinfo{pages}{031003} (\bibinfo{year}{2014}).

\bibitem[{\citenamefont{Nazarov et~al.}(2016)\citenamefont{Nazarov,
  Shulenburger, Morales, and Hood}}]{nazarov2016}
\bibinfo{author}{\bibfnamefont{R.}~\bibnamefont{Nazarov}},
  \bibinfo{author}{\bibfnamefont{L.}~\bibnamefont{Shulenburger}},
  \bibinfo{author}{\bibfnamefont{M.}~\bibnamefont{Morales}}, \bibnamefont{and}
  \bibinfo{author}{\bibfnamefont{R.~Q.} \bibnamefont{Hood}},
  \bibinfo{journal}{Phys. Rev. B} \textbf{\bibinfo{volume}{93}},
  \bibinfo{pages}{094111} (\bibinfo{year}{2016}).

\bibitem[{\citenamefont{Bennett et~al.}(2017)\citenamefont{Bennett, Melton,
  Annaberdiyev, Wang, Shulenburger, and Mitas}}]{Bennett:2017jcp}
\bibinfo{author}{\bibfnamefont{M.~C.} \bibnamefont{Bennett}},
  \bibinfo{author}{\bibfnamefont{C.~A.} \bibnamefont{Melton}},
  \bibinfo{author}{\bibfnamefont{A.}~\bibnamefont{Annaberdiyev}},
  \bibinfo{author}{\bibfnamefont{G.}~\bibnamefont{Wang}},
  \bibinfo{author}{\bibfnamefont{L.}~\bibnamefont{Shulenburger}},
  \bibnamefont{and} \bibinfo{author}{\bibfnamefont{L.}~\bibnamefont{Mitas}},
  \bibinfo{journal}{J. Chem. Phys.} \textbf{\bibinfo{volume}{147}},
  \bibinfo{pages}{224106} (\bibinfo{year}{2017}).

\bibitem[{\citenamefont{Burkatzki et~al.}(2007)\citenamefont{Burkatzki,
  Filippi, and Dolg}}]{Burkatzki:2007jcp}
\bibinfo{author}{\bibfnamefont{M.}~\bibnamefont{Burkatzki}},
  \bibinfo{author}{\bibfnamefont{C.}~\bibnamefont{Filippi}}, \bibnamefont{and}
  \bibinfo{author}{\bibfnamefont{M.}~\bibnamefont{Dolg}}, \bibinfo{journal}{J.
  Chem. Phys.} \textbf{\bibinfo{volume}{126}}, \bibinfo{pages}{234105}
  (\bibinfo{year}{2007}).

\bibitem[{\citenamefont{{I. Ovcharenko, A. Aspuru-Guzik, W. A. {Lester}
  Jr.}}(2001)}]{Lester:2001jcp}
\bibinfo{author}{\bibnamefont{{I. Ovcharenko, A. Aspuru-Guzik, W. A. {Lester}
  Jr.}}}, \bibinfo{journal}{J. Chem. Phys.} \textbf{\bibinfo{volume}{114}},
  \bibinfo{pages}{7790} (\bibinfo{year}{2001}).

\bibitem[{\citenamefont{Pacios and Christiansen}(1985)}]{crenbl}
\bibinfo{author}{\bibfnamefont{L.~F.} \bibnamefont{Pacios}} \bibnamefont{and}
  \bibinfo{author}{\bibfnamefont{P.~A.} \bibnamefont{Christiansen}},
  \bibinfo{journal}{J. Chem. Phys.} \textbf{\bibinfo{volume}{82}},
  \bibinfo{pages}{2664} (\bibinfo{year}{1985}).

\bibitem[{\citenamefont{Stevens et~al.}(1984)\citenamefont{Stevens, Basch, and
  Krauss}}]{SBK}
\bibinfo{author}{\bibfnamefont{W.~J.} \bibnamefont{Stevens}},
  \bibinfo{author}{\bibfnamefont{H.}~\bibnamefont{Basch}}, \bibnamefont{and}
  \bibinfo{author}{\bibfnamefont{M.}~\bibnamefont{Krauss}},
  \bibinfo{journal}{J. Chem. Phys.} \textbf{\bibinfo{volume}{81}},
  \bibinfo{pages}{6026} (\bibinfo{year}{1984}).

\bibitem[{\citenamefont{Dolg et~al.}(1987)\citenamefont{Dolg, Wedig, Stoll, and
  Preuss}}]{STU}
\bibinfo{author}{\bibfnamefont{M.}~\bibnamefont{Dolg}},
  \bibinfo{author}{\bibfnamefont{U.}~\bibnamefont{Wedig}},
  \bibinfo{author}{\bibfnamefont{H.}~\bibnamefont{Stoll}}, \bibnamefont{and}
  \bibinfo{author}{\bibfnamefont{H.}~\bibnamefont{Preuss}},
  \bibinfo{journal}{J. Chem. Phys.} \textbf{\bibinfo{volume}{86}},
  \bibinfo{pages}{866} (\bibinfo{year}{1987}).

\bibitem[{\citenamefont{Trail and Needs}(2005)}]{Trail:2005jcp}
\bibinfo{author}{\bibfnamefont{J.~R.} \bibnamefont{Trail}} \bibnamefont{and}
  \bibinfo{author}{\bibfnamefont{R.~J.} \bibnamefont{Needs}},
  \bibinfo{journal}{J. Chem. Phys.} \textbf{\bibinfo{volume}{122}},
  \bibinfo{pages}{174109} (\bibinfo{year}{2005}),
  \eprint{http://dx.doi.org/10.1063/1.1888569},
  \urlprefix\url{http://dx.doi.org/10.1063/1.1888569}.

\bibitem[{\citenamefont{Dolg}(1996)}]{Dolg:1996cpl}
\bibinfo{author}{\bibfnamefont{M.}~\bibnamefont{Dolg}}, \bibinfo{journal}{Chem.
  Phys. Lett.} \textbf{\bibinfo{volume}{250}}, \bibinfo{pages}{75}
  (\bibinfo{year}{1996}).

\bibitem[{\citenamefont{Reiher and Wolf}(2004)}]{Reiher:2004jcp}
\bibinfo{author}{\bibfnamefont{M.}~\bibnamefont{Reiher}} \bibnamefont{and}
  \bibinfo{author}{\bibfnamefont{A.}~\bibnamefont{Wolf}}, \bibinfo{journal}{J.
  Chem. Phys} \textbf{\bibinfo{volume}{121}}, \bibinfo{pages}{2037}
  (\bibinfo{year}{2004}).

\bibitem[{\citenamefont{Werner et~al.}(2012)\citenamefont{Werner, Knowles,
  Knizia, Manby, and Sch{\"u}tz}}]{MOLPRO-WIREs}
\bibinfo{author}{\bibfnamefont{H.-J.} \bibnamefont{Werner}},
  \bibinfo{author}{\bibfnamefont{P.~J.} \bibnamefont{Knowles}},
  \bibinfo{author}{\bibfnamefont{G.}~\bibnamefont{Knizia}},
  \bibinfo{author}{\bibfnamefont{F.~R.} \bibnamefont{Manby}}, \bibnamefont{and}
  \bibinfo{author}{\bibfnamefont{M.}~\bibnamefont{Sch{\"u}tz}},
  \bibinfo{journal}{WIREs Comput Mol Sci} \textbf{\bibinfo{volume}{2}},
  \bibinfo{pages}{242} (\bibinfo{year}{2012}).

\bibitem[{\citenamefont{{T. H. Dunning Jr.}}(1989)}]{Dunning:1989jcp}
\bibinfo{author}{\bibnamefont{{T. H. Dunning Jr.}}}, \bibinfo{journal}{J. Chem.
  Phys.} \textbf{\bibinfo{volume}{90}}, \bibinfo{pages}{1007}
  (\bibinfo{year}{1989}).

\bibitem[{\citenamefont{Christiansen et~al.}(1979)\citenamefont{Christiansen,
  Lee, and Pitzer}}]{Christiansen:1979jcp}
\bibinfo{author}{\bibfnamefont{P.~A.} \bibnamefont{Christiansen}},
  \bibinfo{author}{\bibfnamefont{Y.~S.} \bibnamefont{Lee}}, \bibnamefont{and}
  \bibinfo{author}{\bibfnamefont{K.~S.} \bibnamefont{Pitzer}},
  \bibinfo{journal}{J. Chem. Phys.} \textbf{\bibinfo{volume}{71}},
  \bibinfo{pages}{4445} (\bibinfo{year}{1979}).

\bibitem[{\citenamefont{Spellucci}(2009)}]{DONLP2}
\bibinfo{author}{\bibfnamefont{P.}~\bibnamefont{Spellucci}},
  \emph{\bibinfo{title}{{DONLP2 Nonlinear Optimization Code}}}
  (\bibinfo{year}{2009}),
  \urlprefix\url{http://www.mathematik.tu-darmstadt.de/fbereiche/numerik/staff/spellucci/DONLP2/}.

\bibitem[{\citenamefont{Spellucci}(1998{\natexlab{a}})}]{SpellucciA}
\bibinfo{author}{\bibfnamefont{P.}~\bibnamefont{Spellucci}},
  \bibinfo{journal}{Math. Prog.} \textbf{\bibinfo{volume}{82}},
  \bibinfo{pages}{413} (\bibinfo{year}{1998}{\natexlab{a}}).

\bibitem[{\citenamefont{Spellucci}(1998{\natexlab{b}})}]{SpellucciB}
\bibinfo{author}{\bibfnamefont{P.}~\bibnamefont{Spellucci}},
  \bibinfo{journal}{Math. Meth. of Oper. Res.} \textbf{\bibinfo{volume}{47}},
  \bibinfo{pages}{355} (\bibinfo{year}{1998}{\natexlab{b}}).

\bibitem[{web(2017)}]{website}
\emph{\bibinfo{title}{http://pseudopotentiallibrary.org}}
  (\bibinfo{year}{2017}), \urlprefix\url{http://pseudopotentiallibrary.org}.

\bibitem[{\citenamefont{{T. H. Dunning Jr., K. A. Peterson, D. E.
  Woon}}(2002)}]{dunningcc}
\bibinfo{author}{\bibnamefont{{T. H. Dunning Jr., K. A. Peterson, D. E.
  Woon}}}, \emph{\bibinfo{title}{Encyclopedia of Computational Chemistry}}
  (\bibinfo{publisher}{Wiley}, \bibinfo{year}{2002}).

\bibitem[{\citenamefont{Kramida et~al.}(2018)\citenamefont{Kramida,
  {Yu.~Ralchenko}, Reader, and {and NIST ASD Team}}}]{NIST_ASD}
\bibinfo{author}{\bibfnamefont{A.}~\bibnamefont{Kramida}},
  \bibinfo{author}{\bibnamefont{{Yu.~Ralchenko}}},
  \bibinfo{author}{\bibfnamefont{J.}~\bibnamefont{Reader}}, \bibnamefont{and}
  \bibinfo{author}{\bibnamefont{{and NIST ASD Team}}},
  \bibinfo{howpublished}{{NIST Atomic Spectra Database (ver. 5.5.6), [Online].
  Available: {\tt{https://physics.nist.gov/asd}} [2018, April 19]. National
  Institute of Standards and Technology, Gaithersburg, MD.}}
  (\bibinfo{year}{2018}).

\bibitem[{\citenamefont{Andersen et~al.}(1999)\citenamefont{Andersen, Haugen,
  and Hotop}}]{EA}
\bibinfo{author}{\bibfnamefont{T.}~\bibnamefont{Andersen}},
  \bibinfo{author}{\bibfnamefont{H.~K.} \bibnamefont{Haugen}},
  \bibnamefont{and} \bibinfo{author}{\bibfnamefont{H.}~\bibnamefont{Hotop}},
  \bibinfo{journal}{Journal of Physical and Chemical Reference Data}
  \textbf{\bibinfo{volume}{28}}, \bibinfo{pages}{1511} (\bibinfo{year}{1999}).

\bibitem[{\citenamefont{Chang et~al.}(1977)\citenamefont{Chang, Habitz, and
  Schwarz}}]{Chang:1977tca}
\bibinfo{author}{\bibfnamefont{T.-C.} \bibnamefont{Chang}},
  \bibinfo{author}{\bibfnamefont{P.}~\bibnamefont{Habitz}}, \bibnamefont{and}
  \bibinfo{author}{\bibfnamefont{W.~H.~E.} \bibnamefont{Schwarz}},
  \bibinfo{journal}{Theoretica chimica acta} \textbf{\bibinfo{volume}{44}},
  \bibinfo{pages}{61} (\bibinfo{year}{1977}).

\bibitem[{\citenamefont{Ranasinghe et~al.}(2015)\citenamefont{Ranasinghe,
  Frisch, and Petersson}}]{ranasinghe2015}
\bibinfo{author}{\bibfnamefont{D.~S.} \bibnamefont{Ranasinghe}},
  \bibinfo{author}{\bibfnamefont{M.~J.} \bibnamefont{Frisch}},
  \bibnamefont{and} \bibinfo{author}{\bibfnamefont{G.~A.}
  \bibnamefont{Petersson}}, \bibinfo{journal}{J. Chem. Phys.}
  \textbf{\bibinfo{volume}{143}}, \bibinfo{pages}{214110}
  (\bibinfo{year}{2015}).

\bibitem[{\citenamefont{Ceperley and Mitas}(1996)}]{CeperleyMitas}
\bibinfo{author}{\bibfnamefont{D.~M.} \bibnamefont{Ceperley}} \bibnamefont{and}
  \bibinfo{author}{\bibfnamefont{L.}~\bibnamefont{Mitas}}, in
  \emph{\bibinfo{booktitle}{Advances in Chemical Physics, Vol. XCIII}}, edited
  by \bibinfo{editor}{\bibfnamefont{I.}~\bibnamefont{Prigogine}}
  \bibnamefont{and} \bibinfo{editor}{\bibfnamefont{S.~A.} \bibnamefont{Rice}}
  (\bibinfo{publisher}{John Wiley and Sons}, \bibinfo{address}{New York},
  \bibinfo{year}{1996}), pp. \bibinfo{pages}{1--38}.

\bibitem[{\citenamefont{Oganov and Dorogokupets}(2003)}]{oganov2003}
\bibinfo{author}{\bibfnamefont{A.~R.} \bibnamefont{Oganov}} \bibnamefont{and}
  \bibinfo{author}{\bibfnamefont{P.~I.} \bibnamefont{Dorogokupets}},
  \bibinfo{journal}{Phys. Rev. B} \textbf{\bibinfo{volume}{67}},
  \bibinfo{pages}{224110} (\bibinfo{year}{2003}).

\bibitem[{\citenamefont{Shirley et~al.}(1991)\citenamefont{Shirley, Mitas, and
  Martin}}]{shirley1991}
\bibinfo{author}{\bibfnamefont{E.}~\bibnamefont{Shirley}},
  \bibinfo{author}{\bibfnamefont{L.}~\bibnamefont{Mitas}}, \bibnamefont{and}
  \bibinfo{author}{\bibfnamefont{R.~M.} \bibnamefont{Martin}},
  \bibinfo{journal}{Phys. Rev. B} \textbf{\bibinfo{volume}{44}},
  \bibinfo{pages}{3396} (\bibinfo{year}{1991}).

\bibitem[{\citenamefont{Shirley and Martin}(1993)}]{martin:prb1993}
\bibinfo{author}{\bibfnamefont{E.~L.} \bibnamefont{Shirley}} \bibnamefont{and}
  \bibinfo{author}{\bibfnamefont{R.~M.} \bibnamefont{Martin}},
  \bibinfo{journal}{Phys. Rev. B} \textbf{\bibinfo{volume}{47}},
  \bibinfo{pages}{15413} (\bibinfo{year}{1993}).

\bibitem[{\citenamefont{Muller et~al.}(1984)\citenamefont{Muller, Flesch, and
  Meyer}}]{muller1984a}
\bibinfo{author}{\bibfnamefont{W.}~\bibnamefont{Muller}},
  \bibinfo{author}{\bibfnamefont{J.}~\bibnamefont{Flesch}}, \bibnamefont{and}
  \bibinfo{author}{\bibfnamefont{W.}~\bibnamefont{Meyer}}, \bibinfo{journal}{J.
  Chem. Phys. Rev.} \textbf{\bibinfo{volume}{80}}, \bibinfo{pages}{329}
  (\bibinfo{year}{1984}).

\bibitem[{\citenamefont{Muller and Meyer}(1984)}]{muller1984b}
\bibinfo{author}{\bibfnamefont{W.}~\bibnamefont{Muller}} \bibnamefont{and}
  \bibinfo{author}{\bibfnamefont{W.}~\bibnamefont{Meyer}}, \bibinfo{journal}{J.
  Chem. Phys.} \textbf{\bibinfo{volume}{80}}, \bibinfo{pages}{3311}
  (\bibinfo{year}{1984}).

\bibitem[{dat(2018)}]{data_doi}
\emph{\bibinfo{title}{{Data for this paper, published in the Materials Data
  Facility }}},
  \bibinfo{howpublished}{{\url{http://dx.doi.org/doi:10.18126/M23S7R } }}
  (\bibinfo{year}{2018}).

\end{thebibliography}

\end{document}